\documentclass[12pt]{article}
\pdfoutput=1
\usepackage[bookmarks=true]{hyperref}
\usepackage{graphicx,epsfig,color}
\usepackage{cite}
\usepackage[english]{babel}
\usepackage{amssymb,amsfonts,amsmath}
\usepackage{verbatim}
\usepackage{tikz}
\usepackage{bbold}
\usepackage{physics}

\newcommand{\be}{\begin{equation}} \newcommand{\ee}{\end{equation}}
\newcommand{\bea}{\begin{eqnarray}} \newcommand{\eea}{\end{eqnarray}}
\newcommand{\beann}{\begin{eqnarray*}}  \newcommand{\eeann}{\end{eqnarray*}}

\newcommand{\p}[0]{\partial}
\newcommand{\bve}{\mbox{{\boldmath$\epsilon$}}}

\usepackage[top=2.5cm, bottom=2.5cm, left=2.5cm, right=2.5cm]{geometry}	
\numberwithin{equation}{section}

\begin{document}

\begin{flushright}
\verb"UTTG-01-2021, HIP-2021-1/TH, BRX-TH-6673"
\end{flushright}

\begin{center}

\centering{\Large {\bf Quantum information probes of charge fractionalization in large-$N$ gauge theories}}

\vspace{6mm}

\renewcommand\thefootnote{\mbox{$\fnsymbol{footnote}$}}
Brandon S. DiNunno,${}^{1}$\footnote{bsd86@physics.utexas.edu}
Niko Jokela,${}^{2,3}$\footnote{niko.jokela@helsinki.fi}
Juan F. Pedraza${}^{4,5}$\footnote{j.pedraza@ucl.ac.uk} and
Arttu P\"onni${}^{6}$\footnote{arttu.ponni@aalto.fi}

\vspace{4mm}

${}^1${\small {\em Theory Group, Department of Physics}\\
\small {\em The University of Texas at Austin, Austin TX 78712, USA}}

\vspace{2mm}
\vskip 0.2cm
${}^2${\small {\em Department of Physics and}} ${}^3${\small {\em Helsinki Institute of Physics}} \\
{\small {\em University of Helsinki, Helsinki FIN-00014, Finland}}

\vspace{2mm}
\vskip 0.2cm
${}^4${\small {\em Department of Physics and Astronomy}\\
\small {\em University College London, London WC1E 6BT, UK}}

\vspace{2mm}
\vskip 0.2cm
${}^5${\small {\em Martin Fisher School of Physics}\\
\small {\em Brandeis University, Waltham MA 02453, USA}}

\vspace{2mm}
\vskip 0.2cm
${}^6${\small {\em Micro and Quantum Systems Group}\\
\small {\em Department of Electronics and Nanoengineering}}\\
{\small {\em Aalto University, Finland}}

\end{center}

\vspace{6mm}

\renewcommand\thefootnote{\mbox{\arabic{footnote}}}

\begin{abstract}
\noindent We study in detail various information theoretic quantities with the intent of distinguishing between different charged sectors in fractionalized states of large-$N$ gauge theories. For concreteness, we focus on a simple holographic $(2+1)$-dimensional strongly coupled electron fluid whose charged states organize themselves into fractionalized and coherent patterns at sufficiently low temperatures. However, we expect that our results are quite generic and applicable to a wide range of systems, including non-holographic. The probes we consider include the entanglement entropy, mutual information, entanglement of purification and the butterfly velocity. The latter turns out to be particularly useful, given the universal connection between momentum and charge diffusion in the vicinity of a black hole horizon. The RT surfaces used to compute the above quantities, though, are largely insensitive to the electric flux in the bulk. To address this deficiency, we propose a generalized entanglement functional that is motivated through the Iyer--Wald formalism, applied to a gravity theory coupled to a $U(1)$ gauge field. We argue that this functional gives rise to a coarse grained measure of entanglement in the boundary theory which is obtained by tracing over (part) of the fractionalized and cohesive charge degrees of freedom. Based on the above, we construct a candidate for an entropic $c$-function that accounts for the existence of bulk charges. We explore some of its general properties and their significance, and discuss how it can be used to efficiently account for charged degrees of freedom across different energy scales.
\end{abstract}

\newpage
\tableofcontents

\section{Introduction}

With the exception of gravity, all fundamental interactions are governed by gauge theories. Understanding the interplay of active degrees of freedom at the quantum level can be notoriously hard, especially when the phases of interest are cold and densely populated. The direst of situations occurs when one lacks a quasiparticle description altogether, making an effective description all the more elusive. Even seemingly standard strongly correlated electron systems continue to source new experimental results that lack a theoretical understanding -- a situation that has endured for decades.

AdS/CFT, or holography, is a framework that has become increasingly popular to understand complicated phenomena involving strongly coupled degrees of freedom in large-$N$ gauge theories \cite{Maldacena:1997re}. Its application to systems with a finite charge density range from neutron stars \cite{deBoer:2009wk,Hoyos:2016zke,Annala:2017tqz,Jokela:2018ers,Hirayama:2019vod,Chesler:2019osn,Ecker:2019xrw,Fadafa:2019euu,Kovensky:2020xif,Demircik:2020jkc,BitaghsirFadafan:2020otb,Jokela:2020piw} to quantum Hall systems \cite{KeskiVakkuri:2008eb,Fujita:2009kw,Alanen:2009cn,Bergman:2010gm,Bayntun:2010nx,Jokela:2010nu,Jokela:2011eb,Lippert:2014jma,Bea:2014yda,Mezzalira:2015vzn,Kristjansen:2016rsc}, superfluids and superconductors \cite{Hartnoll:2008kx,Horowitz:2008bn,Herzog:2009xv,Gauntlett:2009dn,Ammon:2009xh,Horowitz:2010gk,Bhattacharya:2011tra,Bhaseen:2012gg}, strange metals \cite{Hartnoll:2009ns,Sachdev:2010uj,Davison:2013txa,HoyosBadajoz:2010kd,Dong:2012se,Edalati:2012tc,Gath:2012pg,Edalati:2013tma,Pedraza:2018eey}, and more general quantum critical systems \cite{Herzog:2007ij,Faulkner:2009wj,Charmousis:2010zz,Sachdev:2010ch,Gouteraux:2011ce,Hartnoll:2016apf}. Many of these setups exhibit a rich phenomenology that resembles that of real-world (finite-$N$) condensed matter systems. Because of this, holography has become a useful tool in the theorists' model building arsenal, shedding light on a variety of physical phenomena and, oftentimes, uncovering surprising universal properties that are not very sensitive to the details of the model.

In the standard holographic scenario, the dual geometry of a finite density state involves a planar black brane in the bulk, with all $U(1)$ sources cloaked by a horizon. The dual field theory interpretation is that the charges are completely fractionalized and they experience dissipation. However, in the very low temperature regime some of the charges may be located outside the horizon, in which case they are dissipationless and one calls them coherent. If there is a mass gap to such charged excitations, one could simply measure the electrical currents and discern the fractionalized contributions from those of the coherent ones. In the absence of a gap the situation is less clear. In this paper we will devise new probes for charge fractionalization that can be exploited for practical purposes in more generic cases. We note that terminology fractionalized phase used in this paper is not synonymous to a deconfining phase, which also involves the neutral gluon sector.

We will introduce several different probes that are sensitive to the cohesive degrees of freedom. We believe that the lessons learned from this exercise are quite generic and applicable to a wide range of systems. For definiteness, we will illustrate the strength of our analysis through a simple holographic system. Specifically, we will deal with a holographic dual to a $(2+1)$-dimensional strongly coupled electron fluid \cite{Hartnoll:2010gu,Puletti:2010de,Hartnoll:2010ik} which has the property that the charged states organize themselves into a fractionalized and coherent patterns at sufficiently low temperatures.

We will demonstrate that, in addition to electrical conductivities, various information theoretic measures can be used to diagnose whether the active quantum degrees of freedom are coherent or dissipative in the low temperature regime, with entanglement entropy being the prominent example. The computation of the holographic entanglement entropy at large-$N$ and strong coupling is remarkably simple, as it follows from the Ryu--Takayanagi (RT) proposal \cite{Ryu:2006bv,Ryu:2006ef,Nishioka:2009un}. This is quite striking given the fact that in gauge theories even setting up the computation is subtle \cite{Donnelly:2011hn,Casini:2013rba,Radicevic:2014kqa,Donnelly:2014gva,Lin:2018bud}. The physical Hilbert space does not admit a local tensor product decomposition because the physical observables are non-local, see, {\emph{e.g.}}, \cite{Ghosh:2015iwa,Soni:2015yga}. At vanishing density, this problem is circumvented both via classical holographic prescriptions as well as lattice formulations. In the former case one does not need to even invoke bulk gauge fields and in the latter case one carefully avoids making cuts along the links when defining the boundary entangling region upon summing over plaquettes before taking the continuum limit. At finite density, however, the lattice formulation is plagued by the infamous Sign Problem \cite{deForcrand:2010ys}. With so few tools at our disposal, it is therefore interesting to investigate what holographic entanglement entropy can tell us about charge fractionalization.

In addition to entanglement entropy, we explore two other measures of entanglement that are more suitable for characterizing \emph{mixed} quantum states. First we compute the mutual information, a quantity that measures the total amount of correlation between two subsystems (classical and quantum), obeying an area law \cite{GroismanMI,Wolf:2007tdq}. This quantity is constructed from the entanglement entropies of various subregions and so, in a sense, it is not a completely independent measure. It is, however, free of UV divergences, and so it is independent of the way one regularizes the theory. Perhaps more interestingly, we compute the so-called entanglement of purification, which involves an optimized purification of the mixed state \cite{purification} and only measures quantum correlations. Holographically, this quantity has been proposed to be dual to the entanglement wedge cross section\cite{Takayanagi:2017knl,Nguyen:2017yqw}. Unlike mutual information, the entanglement of purification cannot be written solely in terms of entanglement entropies, providing an independent and interesting measure to diagnose bipartite correlations. Finally, we consider a dynamical information-theoretic probe which is related to entanglement entropy in holographic theories: the butterfly velocity \cite{Shenker:2013pqa}. This quantity can be computed by determining the smallest entanglement wedge that contains an infalling bulk perturbation at late times \cite{Mezei:2016wfz}, thus also invoking the same RT surfaces that enter the calculation of entanglement entropy. Under certain assumptions, the butterfly velocity is known to be related to charge diffusion across the horizon \cite{Blake:2016wvh}, though this relation is not expected to hold generically \cite{Davison:2016ngz,Blake:2017qgd}. Nevertheless, we will show that in our setup it can still be a useful probe to help us distinguish between dissipationless degrees of freedom and fractionalized ones.

The RT surfaces that we use to compute the above quantities, though, are insensitive to the electric flux.
The flux forged from the bulk spacetime simply passes through the RT surface with no effect. This is because the RT surface is purely geometric and, therefore, cannot distinguish between the flux emanating from coherent or dissipative charges. An obvious solution would be to allow the extremal surface to ``count'' the flux going through it, or even more explicitly, adjust its shape according to flux contributions. Indeed, the proposal outlined in the work by Hartnoll--Radi\v{c}evi\'c \cite{Hartnoll:2012ux} does exactly this and seems to distinguish between cohesive and fractionalized charges. In this work, we will put the work of \cite{Hartnoll:2012ux} on a more solid footing by proposing a new ``generalized entanglement functional'' $\cal S$ that results from the Iyer--Wald formalism for a gravity theory coupled to a $U(1)$ gauge field. We argue that this quantity can be interpreted in the boundary as a coarser measure of entanglement for
the subsystem, where one traces over (part) of the fractionalized and coherent charge degrees of freedom as one increases the size of the region. As we will show, the generalized functional reduces to the one proposed in \cite{Hartnoll:2012ux} in the IR and gives rise to the needed generalized extremal surfaces in the bulk. Further, it makes contact in the UV with a CFT quantity dubbed ``charged entanglement entropy'' \cite{Belin:2013uta}, which can be verified from the matching of first laws around perturbations of AdS. In general, however, the two quantities differ for general excited states. This mismatch can be traced back to the appearance of a local chemical potential in the generalized functional, which not only measures the flux through the region but also gives it a local weighing.

Armed with the generalized entanglement entropy, we can then ask if we can quantify the number of active charged degrees of freedom at different scales and address whether or not they are dissipationless. To do so, we define a function $\mathcal{C}$ that is built out of generalized entanglement entropies for strip entangling regions, and has all the desired properties for an entropic $c$-function \cite{Casini:2004bw,Nishioka:2006gr,Myers:2010tj,Myers:2012ed,Casini:2012ei,Liu:2012eea,Liu:2013una}. The function $\mathcal{C}$ attains constant values both in the UV and in the IR, values that we derive explicitly and associate with the existence of coherent and fractionalized charges in the bulk. It also decreases monotonically as energy is lowered and hence is a natural candidate for an entropic $c$-function that can be used to diagnose cohesive degrees of freedom in the bulk.

The rest of this paper is organized as follows. In Sec.~\ref{sec:review} we review salient details of the  holographic dual that we use throughout the paper in order to address questions pertaining to charge fractionalization. We continue in Sec.~\ref{sec:info} with a detailed discussion of several probes that reveal useful information about the charged matter at low temperature: the entanglement entropy, mutual information, entanglement of purification and the butterfly velocity. Then, in Sec.~\ref{sec:generalized_functional}, we introduce a new tool which we call the generalized entanglement entropy. We use this new tool to define an entropic $c$-function, $\mathcal{C}$, which counts the amount of bulk charge degrees of freedom across different scales. We conclude in Sec.~\ref{sec:discussion} with a summary of our results and a list of open questions. The paper also contains various appendices detailing intermediate steps in several computations of the main text. App.~\ref{app:vb} contains a discussion of the butterfly velocity. App.~\ref{app:IyerWald} contains the derivation of the generalized entanglement entropy functional; App.~\ref{app:disk} then specializes this functional to the case of disk entangling regions. Finally, App.~\ref{app:cfunctions} contains various analytic limits of the proposed $\mathcal{C}$-function.

\section{Review of electron cloud geometry\label{sec:review}}

In this paper, we will be interested in studying the holographic duals of $(2+1)$-dimensional field theories of strongly interacting fermions at finite temperature and charge density. Therefore, the spacetimes that we will
consider herein are taken to be asymptotically AdS$_4$. At low temperature, the charged AdS$_4$ black hole may undergo a ``brane nucleation'' instability by ejecting its charge to reach an energetically more favorable ground state \cite{Henriksson:2019zph,Henriksson:2019ifu}; in AdS/CFT context this instability has also been called the Fermi seasickness \cite{Hartnoll:2009ns}. In other words, when the backreaction of bulk fermions is taken into account, the AdS$_4$-Reissner--Nordstr{\"o}m black hole is quantum mechanically unstable towards the formation of an electron cloud.
This leads to many interesting physical effects in the boundary theory, some of which have been studied \cite{Hartnoll:2010gu,Hartnoll:2010ik,Puletti:2010de} and more recently in \cite{McInnes:2009zp,Hartnoll:2010xj,Puletti:2011pr,Gran:2018jnt}. More intricate studies including quantum corrections, see, {\emph{e.g.}}, \cite{Medvedyeva:2013rpa,Allais:2013lha}, subsequently confirmed the validity of the electron star (electron cloud) solution even beyond its original range of parameters.

Let us now be more specific about the setup used in the present paper. We will consider systems with charged fermions in the bulk modeled as ideal fluids, namely the electron cloud solution \cite{Hartnoll:2010ik,Puletti:2010de} that constitutes the finite temperature generalization of the electron star \cite{Hartnoll:2010gu, Hartnoll:2010xj}. After studying some basic thermodynamic quantities of the system with a view towards condensed matter applications, we proceed to investigate how the charge is distributed in the geometry using tools familiar from quantum information theory. Let us thus start by reviewing and collecting some useful facts about the electron cloud solution. The Einstein--Maxwell theory with a negative cosmological constant and a charged perfect fluid component has the action
\be
 S = \int d^4x\sqrt{|g|} \left( \mathcal L_E + \mathcal L_{EM} + \mathcal L_{fluid} \right)  \ ,
\ee
with Lagrangians
\bea
\mathcal L_E & = & \frac{1}{2\kappa^2} \left( R + \frac{6}{L^2} \right) \ ,\\
  \mathcal L_{EM} & = & -\frac{1}{4 e^2} F_{\mu\nu} F^{\mu\nu} \ ,\\
  \mathcal L_{fluid} & = & -\rho(\sigma) + \sigma u^\mu \left( \partial_\mu \phi + A_\mu \right) + \lambda (u^\mu u_\mu + 1) \ ,
\eea
where $\kappa^2=8\pi G_N$, $u^\mu$, $\rho$, and $\sigma$ are the velocity, energy density, and charge density of the fluid, and $\phi,\,\lambda$ are auxiliary fields which we will put on-shell. The resulting equations of motion read\cite{Hartnoll:2010gu}
\be\label{eq:EMF}
R_{\mu \nu}- \frac{1}{2}g_{\mu \nu} R - \frac{3}{L^2}g_{\mu \nu} = \kappa^2 \left( T^{\mathrm{EM}}_{\mu \nu}+T^{\mathrm{fluid}}_{\mu \nu}\right), \qquad \nabla^{\nu} F_{\mu \nu} = e^2 J^{\mathrm{fluid}}_{\mu} \ ,
\ee
where the sources are given by
\bea
T^{\mathrm{EM}}_{\mu \nu} & = & \frac{1}{e^2}\left( F_{\mu \lambda} F_{\nu}^{\lambda}- \frac{1}{4} g_{\mu \nu} F_{\lambda \sigma}F^{\lambda \sigma}\right) \ , \\
T^{\mathrm{fluid}}_{\mu \nu} & = &\left( \rho + p \right) u_{\mu}u_{\nu} + p g_{\mu \nu} \ ,\\
 J^{\mathrm{fluid}}_{\mu} & = & \sigma u_{\mu}\ .
\eea
In addition, we have the constraint $u^\mu u_\mu = -1$. The Ansatz for a static, planar black brane metric, and a Maxwell EM field is chosen as
\be\label{eq:metric}
ds^2 =-f(v) dt^2 + \frac{1}{v^2}\left(dx_1^2+dx_2^2 \right) + g(v) dv^2, \qquad A= A_t dt = \frac{e}{\kappa}h(v) dt \ .
\ee
Here and below we have set the AdS radius to unity, $L=1$, but it can be easily restored via dimensional analysis whenever necessary. In these coordinates, $v$ approaches zero at the boundary and the horizon is located at a finite radial position $v_H$. In the absence of a black hole in the bulk, corresponding to setting the temperature to zero ($T=0$), the Poincar\'e horizon is at $v=\infty$.

Our Ansatz is invariant under the scaling:
\begin{equation}
    (t,x,y,v) \rightarrow (t,x,y,v)/\xi\ , \quad f\rightarrow \xi^2  f \ , \quad  g\rightarrow \xi^2 g\ , \quad  h\rightarrow \xi  h \ ,
\end{equation}
and so, assuming the presence of a horizon, we can rescale all quantities by the horizon radius $v_H$ and replace them with their dimensionless counterparts, which we decorate with hats.\footnote{In the hatted variables, the radial coordinate $\hat v\equiv v/v_H$ runs from 0 (at the boundary) to 1 (at the horizon).
} For the fluid variables, we have:
\be
 \hat{p}=\kappa^2 p\ , \quad \hat{\rho}=\kappa^2 \rho \ , \quad \hat{\sigma}= e \kappa \sigma\ .
\ee
We note that as $T\rightarrow 0$, another (Lifshitz) scaling symmetry emerges, which we will comment on in Subsection~\ref{sec:thermo}.

Now we can express the equations of motion \eqref{eq:EMF} as
\begin{align}
\frac{\hat{f}'(\hat{v})}{\hat{v}\hat{f}(\hat{v})}-\frac{\hat{h}'(\hat{v})^2}{2 \hat{f}(\hat{v})} + \hat{g}(\hat{v}) \left( 3 + \hat{p}(\hat{v})\right) -\frac{1}{\hat{v}^2} &=0 \ , \label{eq:EMFdimless}\\
\hat{h}''(\hat{v})+\left( \frac{\hat{v}\hat{h}(\hat{v}) \hat{h}'(\hat{v})}{2\sqrt{\hat{f}(\hat{v})}}  -  \sqrt{\hat{f}(\hat{v})} \right)\hat{g}(\hat{v}) \hat{\sigma}(\hat{v}) &= 0 \ ,\\
\frac{1}{\hat{v}}\left( \frac{\hat{f}'(\hat{v})}{\hat{f}(\hat{v})}+ \frac{\hat{g}'(\hat{v})}{\hat{g}(\hat{v})}+ \frac{4}{\hat{v}}\right)+  \frac{\hat{h}(\hat{v})}{\sqrt{\hat{f}(\hat{v})}} \hat{g}(\hat{v}) \hat{\sigma}(\hat{v}) &= 0 \ . \label{eq:EMFdimless2}
\end{align}
Here and subsequently, the primes will indicate derivation with respect to $\hat{v}$. The equations of motion are further seen to imply a radially conserved current,
\begin{equation}
  J_\xi = \frac{2  \hat v^2  \hat h(\hat v)  \hat h'(\hat v) - ( \hat v^2  \hat f(\hat v))'}{ \hat v^4 \sqrt{ \hat f(\hat v)  \hat g(\hat v)}} \ ,
\end{equation}
with $\partial_{v}  J_\xi = 0$.

The interesting regime for this construction turns out to be a region of parameter space for which it is consistent to assume:
\begin{itemize}
\item A locally flat space approximation in which the fermion physics is correctly captured by an effective local chemical potential, given by
\be\label{eq:muloc}
 u^{\hat t}\hat A_{\hat t}(\hat v) \equiv \hat\mu(\hat v)_{loc} \equiv \frac{\hat{h}(\hat{v})}{\sqrt{\hat{f}(\hat{v})}} \ .
\ee
From now on, to suppress unnecessary notation, we will simply write $\hat\mu(\hat v)_{loc} \to \hat\mu_{loc}$, where the subscript `loc' reminds the reader that this is not a chemical potential of the boundary theory but merely the value of the gauge potential (in the tangent frame) at a given radial position $\hat v$. In addition, we assume that the fermions are cold with equation of state: 
\begin{equation}
  -\hat{p}= \hat{\rho}-\hat{\mu}_{loc}\, \hat{\sigma} \ ,
\end{equation}
where:
\begin{equation}
\hat{\rho} =  \hat{\beta} \int_{\hat{m}}^{\hat{\mu}_{loc}} d \epsilon \; \epsilon^2 \sqrt{\epsilon^2-\hat{m}^2} \ , \qquad \hat{\sigma} =  \hat{\beta} \int_{\hat{m}}^{\hat{\mu}_{loc}} d \epsilon \; \epsilon \sqrt{\epsilon^2-\hat{m}^2} \ ,
\end{equation}
and where the dimensionless constants are $\hat{\beta}=\frac{e^4}{\kappa^2}\beta$ and $\hat{m}^2=\frac{e^2}{\kappa^2}m^2$. This approximation is valid when the Compton wavelength of the fermions is small compared to the radius of curvature.
\item A classical bulk geometry with an order one backreaction of the fermion fluid. This happens when the source terms of the Einstein equations are sizable. For the fermion fluid contributions this can be expressed as
\be
  \hat\beta \sim 1 \ , \qquad \hat{m}^2 \sim 1 \ .
\ee
\end{itemize}
The equations of motion \eqref{eq:EMFdimless}-\eqref{eq:EMFdimless2} admit a charged, planar AdS$_4$-RN black brane solution for the vacuum $\hat{\rho}=\hat{p}=\hat{\sigma}=0$:
\be
 \hat{f}(\hat{v})= \frac{1}{\hat{v}^2}+\frac{\hat q^2}{2}\hat{v}^2 -\left(1+\frac{\hat q^2}{2} \right)\hat{v} \ , \qquad \hat{g}(\hat{v})= \frac{1}{\hat{v}^4 \hat{f}(\hat{v})} \ , \ \hat{h}(\hat{v}) = \hat{q}(1 - \hat{v}) \ .
\ee
Here, the time coordinate has been rescaled to fix the overall normalization of $\hat{f}(\hat{v})$. The dimensionless constant $\hat{q}$ is related to the charge of the black brane. Provided $\hat{q}^2<6$, the AdS$_4$-RN black brane is non-extremal and $\hat{\mu}_{loc}=0$ at the (non-degenerate) horizon. The local chemical potential grows away from the horizon, but only when
\be\label{eq:mucond}
 \hat{\mu}^2_{loc} > \hat{m}^2
\ee
is satisfied, can the fermion fluid be supported.\footnote{This can be rephrased as $\hat{q}^2 > r(\hat{v})$ with the function $r(\hat{v})<6$ as defined in \cite{Puletti:2010de}, eq. (2.16), where the conditions for the existence of a massive fermion fluid are discussed in more detail. There it was shown that for $\hat{m}^2<1$ and $\hat{q}^2<6$, the fermion fluid exists for a finite range $\hat{v}_o>\hat{v}>\hat{v}_i$. The endpoints correspond to the inner and outer edges of the electron cloud. Above some critical temperature, there is only a black brane without a fermion fluid in the bulk.}
After fixing the parameters $\hat{\beta}, \hat{m}$, and $\hat{q}^2$ in the allowed range, one proceeds to numerically integrate eqs. \eqref{eq:EMFdimless}-\eqref{eq:EMFdimless2} inside the electron cloud. To this end, one imposes initial values for $\hat{f}(\hat{v}=\hat{v}_i),\hat{g}(\hat{v}=\hat{v}_i)$, and $\hat{h}(\hat{v}=\hat{v}_i)$ at the inner edge of the cloud. The numerical integration stops at some $\hat{v}_s < \hat{v}_o$, where the condition \eqref{eq:mucond} ceases to be satisfied.

The final step in the construction is to match the numerical solution onto a charged, planar AdS$_4$-RN black brane solution at $\hat{v}=\hat{v}_s$ to yield the exterior solution
\be
 \hat{f}(\hat{v})= c_s^2 \hat{v}^{-2}+ \frac{q_s^2}{2}\hat{v}^{2}- m_s \hat{v}, \qquad \hat{g}(\hat{v})=\frac{c_s^2}{\hat{v}^4 \hat{f}(\hat{v})}, \quad \hat{h}(\hat{v})=\mu_s - q_s \hat{v} \ ,\label{eq:exterior_soluion}
\ee
where
\bea
 c_s^2 & = & \hat{f}(\hat{v}_s)\hat{g}(\hat{v}_s) \hat{v}_s^{4} \ ,\\
 q_s & = & - \hat{h}'(\hat{v}_s) \ ,\\
 \mu_s & = & \hat{h}(\hat{v}_s)- \hat{v}_s \hat{h}'(\hat{v}_s) \ ,\\
 m_s & = & \hat{f}(\hat{v}_s)\hat{g}(\hat{v}_s) \hat{v}_s +\frac{1}{2} \hat{h}'(\hat{v}_s)^2 \hat{v}_s -\hat{f}(\hat{v}_s)\hat{v}_s^{-1}\ .
\eea
Notice that the parameter $\hat q$ corresponding to the charge of the inner RN black hole is an input parameter, while the physical quantity is the chemical potential of the boundary theory, extracted via
\be\label{relationqhat}
 \frac{T}{\mu}  = \frac{6-\hat q^2}{8\pi \mu_s} \ .
\ee

\begin{figure}[t!]
 \centering
     \includegraphics[width=0.3\textwidth]{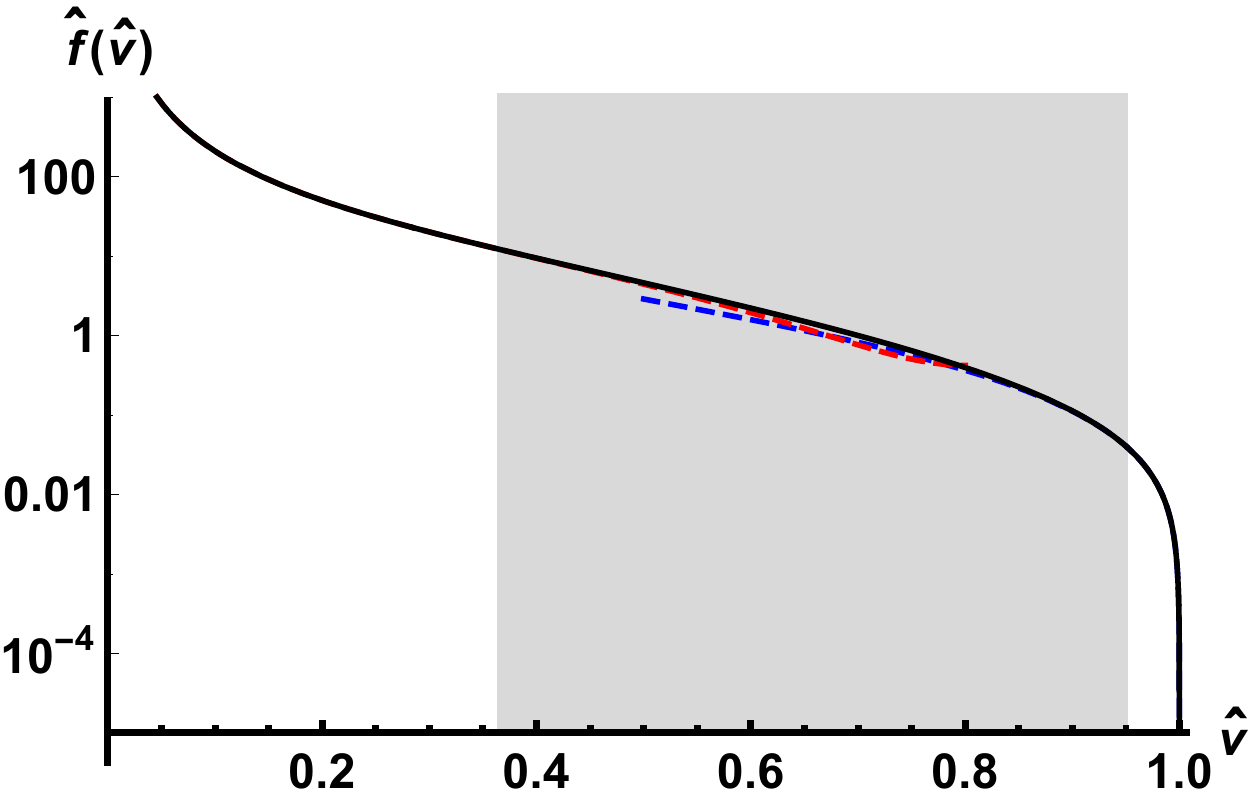}
     \includegraphics[width=0.3\textwidth]{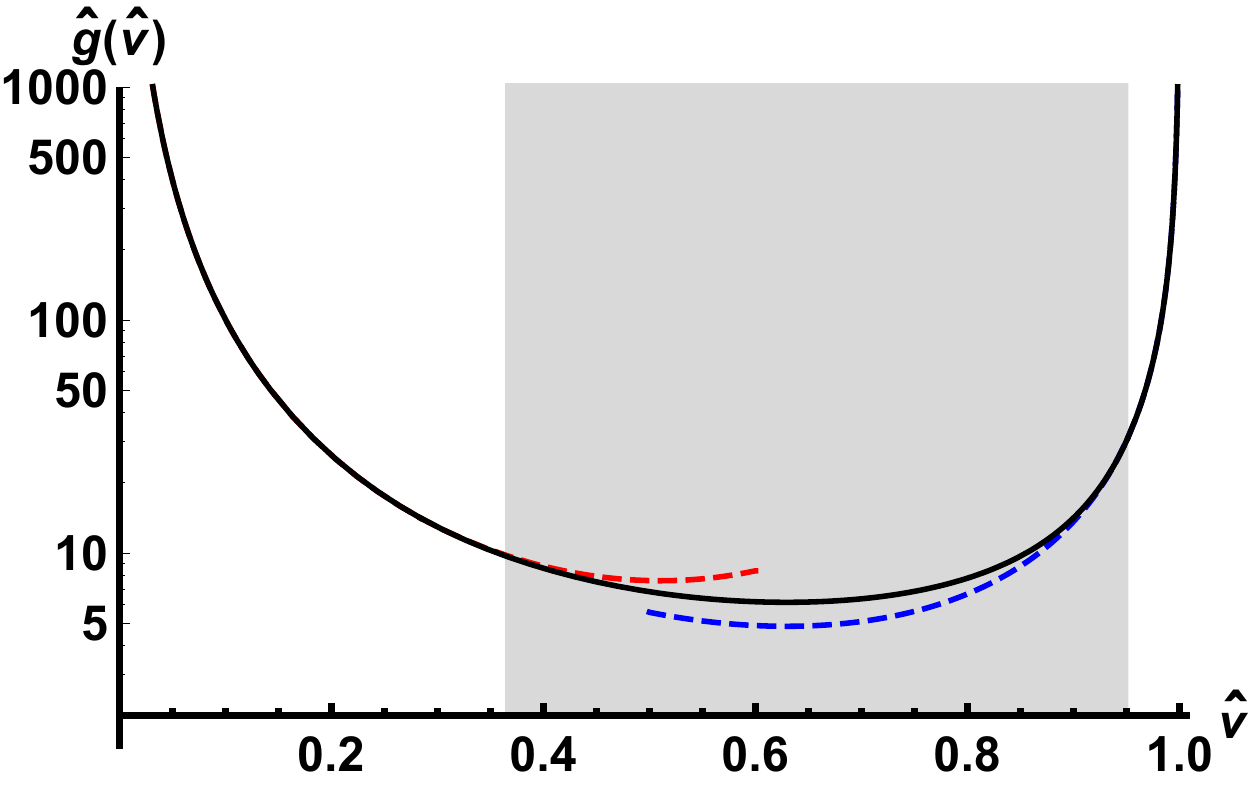}
     \includegraphics[width=0.3\textwidth]{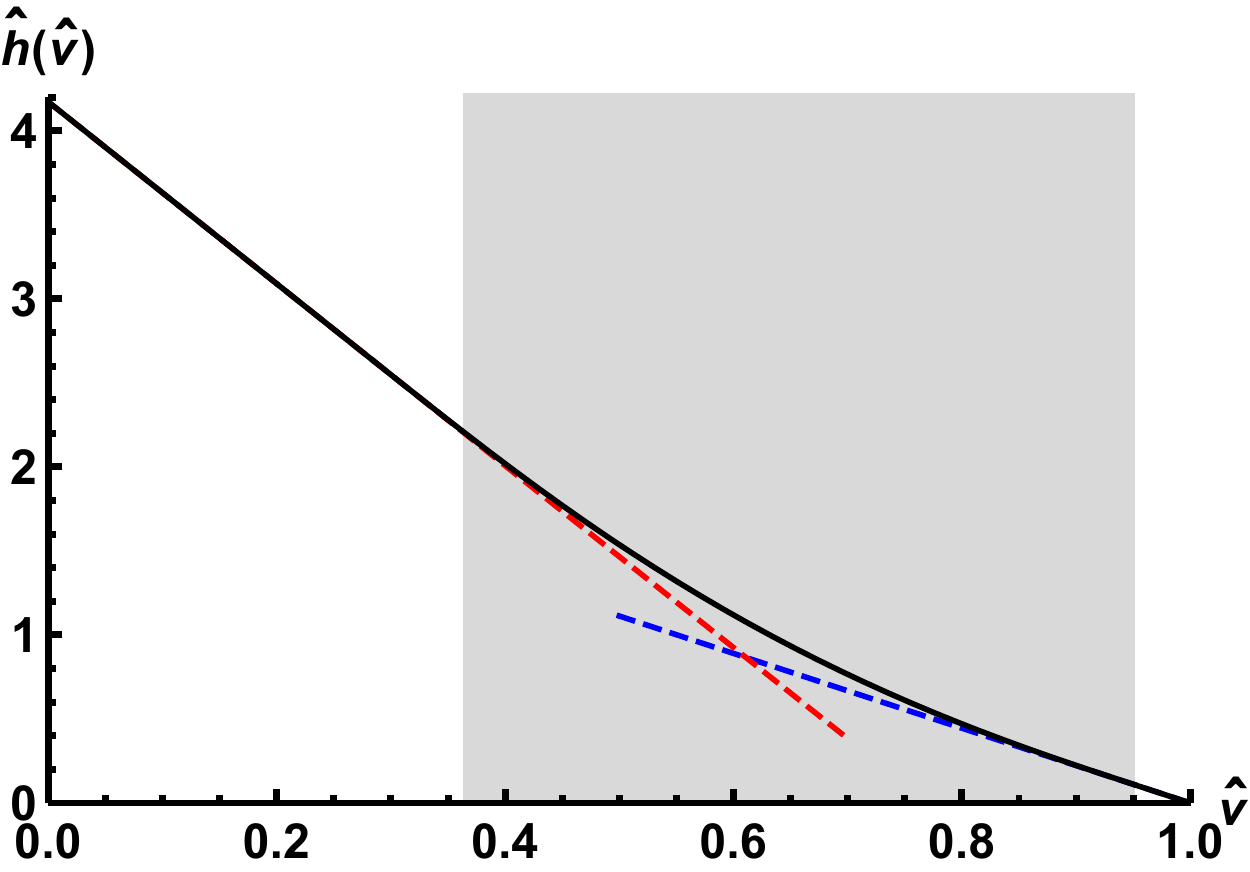}
 \caption{\small The numerical results for the functions $\hat f(\hat v)$, $\hat g(\hat v)$, and $\hat h(\hat v)$ at low temperature $T/\mu=0.01$ and for $\hat m=0.55,\hat\beta=10$. The shaded  gray denotes the bulk region where the cloud resides. Outside of this region the geometry is smoothly attached to two RN black branes, with the dashed blue and red curves corresponding to the inner and outer RN geometries.}
   \label{fig:ECfunctions}
\end{figure}

It turns out to be useful to work with the following set of equations,\footnote{As a consistency check on our numerics, we have used these equations to confirm the various numerical results obtained in \cite{Puletti:2010de}.} which makes some of the physics more transparent:
\bea
 \hat p'(\hat{v}) - \hat\sigma(\hat{v})\frac{d}{d\hat v}\left(\frac{\hat h(\hat{v})}{\sqrt{\hat f(\hat{v})}}\right) &  = & 0 \ ,\label{eq:EC_H1} \\
\frac{d}{d\hat v}\left(\frac{\hat h'(\hat{v})}{\hat v^2\sqrt{\hat f(\hat{v}) \hat g(\hat{v})}}\right) - \frac{\sqrt{\hat g(\hat{v})}}{\hat v^2}\hat\sigma(\hat{v}) & = & 0 \ , \label{eq:EC_H4}\\
    \frac{d}{d\hat v}\log(\hat f(\hat{v}) \hat g(\hat{v}) \hat v^4)+\hat v\frac{\hat h(\hat{v})}{\sqrt{\hat f(\hat{v})}}\hat g(\hat{v})\hat \sigma (\hat{v}) & = & 0 \ , \label{eq:EC_H2} \\
    \frac{1}{\hat v}\left(\frac{\hat f'(\hat{v})}{\hat f(\hat{v})}-\frac{1}{\hat v}\right)-\frac{\hat h'(\hat{v})^2}{2 \hat f(\hat{v})}+\hat g(\hat{v})(\hat p(\hat{v})+3) & = & 0 \ . \label{eq:EC_H3}
\eea
The first equation above is the Gibbs-Duhem relation, a thermodynamic identity at vanishing temperature, and \eqref{eq:EC_H4} is Gauss' law.
Lastly we point out that the above system of equations give us the following expressions for the fluid variables:
\bea
\hat p(\hat v)&=& -3 - \frac{\hat f'(\hat v)}{\hat v \hat f(\hat v)\hat g(\hat v)}+\frac{\hat h'(\hat v)^2}{2\hat f(\hat v)\hat g(\hat v)}+\frac{1}{\hat v^2 \hat g(\hat v)} \ ,\\
\hat \rho(\hat v) &=& 3-\frac{\hat g'(\hat v)}{\hat v \hat g(\hat v)^2}-\frac{\hat h'(\hat v)^2}{2 \hat f(\hat v)\hat g(\hat v)}-\frac{5}{\hat v^2 \hat g(\hat v) } \ ,\\
\hat \sigma(\hat v)&=& \frac{\hat v^2}{\sqrt{\hat g(\hat v)}} \frac{d}{d\hat v} \left( \frac{\hat h'(\hat v)}{\hat v^2 \sqrt{\hat f(\hat v) \hat g(\hat v)}} \right) \ . \label{eq:sigma}
\eea
For reference, the functions $\hat f(\hat v)$, $\hat g(\hat v)$, and $\hat h(\hat v)$ are plotted in Fig.~\ref{fig:ECfunctions} for representative values of the electron cloud where both of the solutions co-exist.

\subsection{Comments on the thermodynamics of the electron cloud}\label{sec:thermo}

Here we want to summarize and expand on previous work relating to the thermodynamics of the electron cloud solution \cite{Hartnoll:2010ik, Puletti:2010de}. The key features of the electron star/cloud solutions are the following: Firstly, at $T=0$, they provide a holographic framework for metallic quantum criticality, {\emph{i.e.}}, at low energies the electron cloud features emergent Lifshitz scaling with a finite dynamical critical exponent,
\bea
 f(v \to \infty) \sim v^{-2z} \ , \quad g(v \to \infty) \sim g_{\mathrm{IR}} v^{-2} \ , \quad h(v\to \infty) \sim h_{\mathrm{IR}} v^{-z} \ ,
\eea
where
\bea
 g_{\mathrm{IR}}^2 = \frac{36 (z-1)z^4}{((1-\hat m^2)z-1)^3 \hat\beta^2} \  , \qquad h_{\mathrm{IR}}^2 = \frac{z-1}{z} \ .
\eea
Secondly, the existence of a smeared Fermi surface has some interesting physical consequences \cite{Hartnoll:2010xj, Puletti:2011pr}. \\
Recall that the local thermodynamics of the charged fermion fluid is determined by the `local' chemical potential $ \hat \mu_{loc} = \hat h(\hat v) /{\scriptstyle \sqrt{\hat f(\hat v)}}$. An important result of \cite{Hartnoll:2010ik, Puletti:2010de} revealed that there is a phase transition between the electron cloud and the black brane solutions, for fixed chemical potential. Namely, the difference in (dimensionless) free energies is
\bea
\Delta \left(\frac{\Omega}{\mu^3}\right) = \left(\frac{\Omega}{\mu^3}\right)_{\mathrm{AdS-RN}} - \left(\frac{\Omega}{\mu^3}\right)_{\mathrm{EC}} \sim \left( \frac{(T_c - T)}{\mu} \right)^3,
\eea
and thus the electron cloud undergoes a third order phase transition to collapse to an AdS$_4$-Reissner--Nordstr\"om black brane above a critical temperature $T_c$ determined by the chemical potential and the mass of the fermions.

Another thermodynamic property of interest is the entropy density of the cloud. It can be shown via dimensional analysis that, at low temperatures $T \ll \mu$, the entropy density $s_{\text{th}}$ scales as
\bea\label{eq:entropy}
s_{\text{th}}= - \left( \frac{\partial \Omega}{\partial T } \right)_{\mu} \sim T^{2/z} \ , \  \frac{T}{\mu}\to 0 \ ,
\eea
in terms of the critical dynamical exponent $z=z(\beta, m)$. This expectation can also be confirmed numerically, as was demonstrated in \cite{Hartnoll:2010ik}. This peculiar scaling with $T$ is in stark contrast with the expected result for an AdS$_4$-Reissner--Nordstr\"om black brane in the small temperature limit. In the latter case, the entropy remains finite as $T\to0$, $s_{\text{th}}\sim \mu^{2}$, signaling a highly degenerate ground state. At high temperature, on the other hand,  the electron cloud ceases to exist and the entropy of the system behaves in the standard way, $s_{\text{th}}\sim T^2$,
which follows from conformal invariance in the UV.

The specific heat capacity $C_v$ is a physical quantity of matter from which useful information, {\emph{e.g.}},  about the nature of quasiparticle excitations can be gleaned. For example, while the specific heat of a fermion liquid exhibits a linear behavior, a bosonic gas scales as $C_v\sim T^2$ (in 2+1 dimensions).
Using (\ref{eq:entropy}), we find that for low temperatures,
\bea
C_v= T \left( \frac{\partial s_{\text{th}}}{\partial T}\right)_{\rho, V} \sim T ^{2/z}  \ , \  \frac{T}{\mu}\to 0 \ ,
\eea
which implies that we recover the result expected for a boson gas in the limit $z\to1$. This is the so-called massless limit, $\hat m=0, \hat\beta \to \infty$, considered in \cite{Hartnoll:2010ik}. Notice that for judicious choices of $\hat m,\hat\beta$ \cite{Hartnoll:2010gu} result in $z=2$ at the IR, a peculiar linear heat capacity associated with strange metals. Furthermore, the speed of first/normal sound can be obtained as the derivative of the pressure with respect to the mass/energy density. In the grand canonical ensemble, the pressure $p=-\Omega$. For the massless case, the conservation of the dilatation current is ensured by a conformal Ward identity which requires $ p = \frac{z}{d} \epsilon$. Thus,
\be
 c_s^2 = \left(\frac{\partial p}{\partial \epsilon}\right)_{\mu} = \frac{z}{d} \ , \ \frac{T}{\mu}\to 0 \ ,
\ee
where $d$ denotes the number of spatial dimensions. However, in general, we would have to compute the speed of first sound numerically. It would be particularly interesting to see if it would result in stiff phases whose $c_s^2$ is above the conformal value \cite{Hoyos:2016cob,Ecker:2017fyh}. This phenomenon can be associated with short-range repulsive interactions \cite{Hoyos:2019kzt,Hoyos:2020fjx}, which, given the fermionic nature of the bulk excitations, is expected at low temperature. Finally, we would like to point out that an analysis of the QNM spectra appeared in \cite{Gran:2018jnt}. It would be an interesting extension thereof to make a connection between the zero sound studied there, the normal sound above, and the butterfly velocity (studied in the next section) in the massless limit.

\section{Information theoretic probes of fractionalized states\label{sec:info}}

Entanglement is an essential feature of quantum mechanics with no classical counterpart. In pure quantum states, the amount of entanglement is uniquely characterized by entanglement entropy. In general QFTs, entanglement entropy is a difficult quantity to compute. This is not the case, however, in holographic systems, where one can simply use the Ryu--Takayanagi formula \cite{Ryu:2006bv,Ryu:2006ef,Nishioka:2009un} which gives the entanglement entropy in terms of the area of a certain bulk extremal surface.

In mixed quantum states the situation is more complicated. There are many interesting, inequivalent measures of quantum and/or classical correlations, and only a few have well-established holographic duals. One simple quantity that we can readily compute is the mutual information \cite{GroismanMI,Wolf:2007tdq}. Mutual information measures the \emph{total} amount of correlation, both classical and quantum mechanical, between given subsystems. It can be defined in terms of entanglement entropy, which makes it an easy quantity to compute in AdS/CFT.
Another interesting quantity to compute is the entanglement of purification \cite{purification}, which measures the amount of \emph{quantum} correlations for a specific ``optimal'' purification. There is a proposal for the holographic dual of the entanglement of purification, called the entanglement wedge cross section \cite{Takayanagi:2017knl,Nguyen:2017yqw}. Unlike mutual information, the entanglement of purification requires further input besides entanglement entropy and thus it is an interesting quantity to study holographically. Both, mutual information and entanglement of purification have been successfully used to characterize strongly coupled phases of matter, including finite density states and quantum critical systems \cite{Kundu:2016dyk,Jokela:2019ebz,Liu:2019qje,BabaeiVelni:2019pkw,Huang:2019zph,Jokela:2019tsb,Ebrahim:2020qif,Amrahi:2020jqg,BabaeiVelni:2020wfl,Fu:2020oep,Lala:2020lcp,Gong:2020pse,Jain:2020rbb,Khoeini-Moghaddam:2020ymm,Saha:2021kwq,Amrahi:2021lgh}.

Finally, we will also consider a dynamical probe of the cloud, often discussed in the quantum information theory context, which serves as a diagnostic of many-body quantum chaos: the butterfly velocity \cite{Shenker:2013pqa}. This quantity measures how quickly the systems reacts to arbitrary local perturbations and can be computed by determining the smallest entanglement wedge that contains an infalling bulk perturbation at late times \cite{Mezei:2016wfz}. Interestingly, there is a known connection between the butterfly velocity and transport across black hole horizons \cite{Blake:2016wvh,Davison:2016ngz,Blake:2017qgd} which, as we show below, prove useful for diagnosing existence of dissipationless charged degrees of freedom.

\subsection{Entanglement entropy}

For a bipartite quantum system described by a density matrix $\rho_{AB}$, the entanglement entropy of a subsystem $A$ is defined as the von Neumann entropy associated with its reduced density matrix $\rho_A = \Tr_B \rho_{AB}$, {\emph{i.e.}},
\begin{align}
  S(A) = -\Tr(\rho_A \log \rho_A) \ .
\end{align}
In AdS/CFT, the entanglement entropy in the Einstein frame is given by \cite{Ryu:2006bv}
\begin{align}
  S(A) = \frac{1}{4 G_N} \text{Area}(\Gamma_A) \ ,
\end{align}
where $\Gamma_A$ is the minimal area, codimension-2 bulk surface lying on a space-like slice,\footnote{Our background is static so all RT surfaces can be taken to be on a canonical time slice $t=$ constant.} which is anchored on the boundary of the entanglement surface $\partial \Gamma_A = \partial A$ and is homologous to $A$. Recall also the relationship $\kappa^2=8\pi G_N$.

We will consider a strip as our entangling region, $A=\{(x,y)|-l/2\leq x \leq l/2, -L_y/2\leq y \leq L_y/2\}$, where $l$ is the width of the strip along the $x$-direction and we consider the limit $L_y\to\infty$. Due to the symmetries of the background and the infinite extent of the strip in the $y$-direction, the profile of the strip can be represented with a single function $\hat{x}=\hat{x}(\hat{v})$. The entanglement entropy then becomes
\begin{align}
  S(l) = \frac{L_y}{4 G_N v_H} \int \frac{d\hat v}{\hat v^2} \sqrt{\hat v^2 \hat g(\hat v) + \hat x'(\hat v)^2} \ .
\end{align}
Since the functional does not depend explicitly on $\hat x(\hat v)$, there is an associated conserved quantity along the surface:
\begin{align}\label{eq:conserv}
  \frac{\hat x'(\hat v)}{\hat v^2\sqrt{\hat v^2 \hat g(\hat v) + \hat x'(\hat v)}} = -\frac{1}{\hat v_*^2} \ .
\end{align}
The integration constant $\hat v_*$ gives the \emph{turning point}, {\emph{i.e.}}, the point where the minimal surface reaches deepest into the bulk. At the turning point, the profile is completely flat, so the first derivative diverges $\hat x'(\hat v_*)\to-\infty$. This fact, together with the conservation equation (\ref{eq:conserv}) can be used to solve for the first derivative of the profile
\begin{align}\label{eq:profile}
  \hat x'(\hat v) = \pm \frac{\hat v^3 \sqrt{\hat g(\hat v)}}{\sqrt{\hat v_*^4 - \hat v^4}} \ .
\end{align}
Finally, equation (\ref{eq:profile}) can be used to express the length of the strip $l$ and entanglement entropy $S(l)$ as follows:
\begin{align}
  \frac{l(\hat v_*)}{v_H} &= 2 \int_0^{\hat v_*} \frac{\hat v^3 \sqrt{\hat g(\hat v)}}{\sqrt{\hat v_*^4 - \hat v^4}} d\hat v \,,\label{eq:strip_width} \\
  \frac{4 G_N v_H S(\hat v_*)}{L_y} &= \frac{2}{\hat\epsilon} - \frac{2}{\hat v_*} + 2\int_0^{\hat v_*} \left(\frac{\sqrt{\hat g(\hat v)}}{\hat v\sqrt{1-\frac{\hat v^4}{\hat v_*^4}}} - \frac{1}{\hat v^2}\right) d\hat v \ , \label{eq:strip_entropy}
\end{align}
where $\hat\epsilon$ is the UV-cutoff in $\hat v$. Above we have written the area law divergence explicitly such that the remaining integrals are convergent. From now on, however, we will consider the \emph{regularized} entropy which we define as the above formula with the $2/\hat\epsilon$-term subtracted.

Before proceeding further, we note that in the above formulas the strip's length $l$ is always accompanied by a factor of $1/v_H$. Hence, it will be useful to
interpret this scale in terms of field theory variables, and study how it appears in the different regimes of interest of entanglement entropy. Following \cite{Kundu:2016dyk,Kundu:2016cgh}, we interpret this scale as an \emph{effective temperature},
\be
T_{\text{eff}}(T,\mu)=\frac{3}{4\pi v_H}\,.
\ee
To understand this interpretation, we note that the horizon's area scale as $A\propto 1/v_H^2$, hence the thermal entropy of the state follows a Stefan-Boltzmann law at temperature $T_{\text{eff}}$ for all $T/\mu$. For instance, in the AdS$_4$-RN black brane the above quantity has the property that
\be
\left(T_{\text{eff}}\right)_{\mathrm{AdS-RN}}\sim T\quad\,\, (T/\mu\to\infty)\,,\qquad \,\, \left(T_{\text{eff}}\right)_{\mathrm{AdS-RN}}\sim \mu\quad\,\, (T/\mu\to0)\,.
\ee
In the electron cloud system, however, we have that
\be
\left(T_{\text{eff}}\right)_{\mathrm{EC}}\sim T\quad\,\, (T/\mu\to\infty)\,,\qquad \,\, \left(T_{\text{eff}}\right)_{\mathrm{EC}}\sim T^{1/z}\mu^{1-1/z}\quad\,\, (T/\mu\to0)\,,
\ee
reflecting the new scaling behavior in the IR. These scaling limits can be easily verified from our numerics. The dependence with $T$ and $\mu$ in the EC can also be deduced from the temperature dependence of the thermal entropy (\ref{eq:entropy}) and dimensional analysis.

\begin{figure}[t!]
  \centering
  \includegraphics[width=1.0\textwidth]{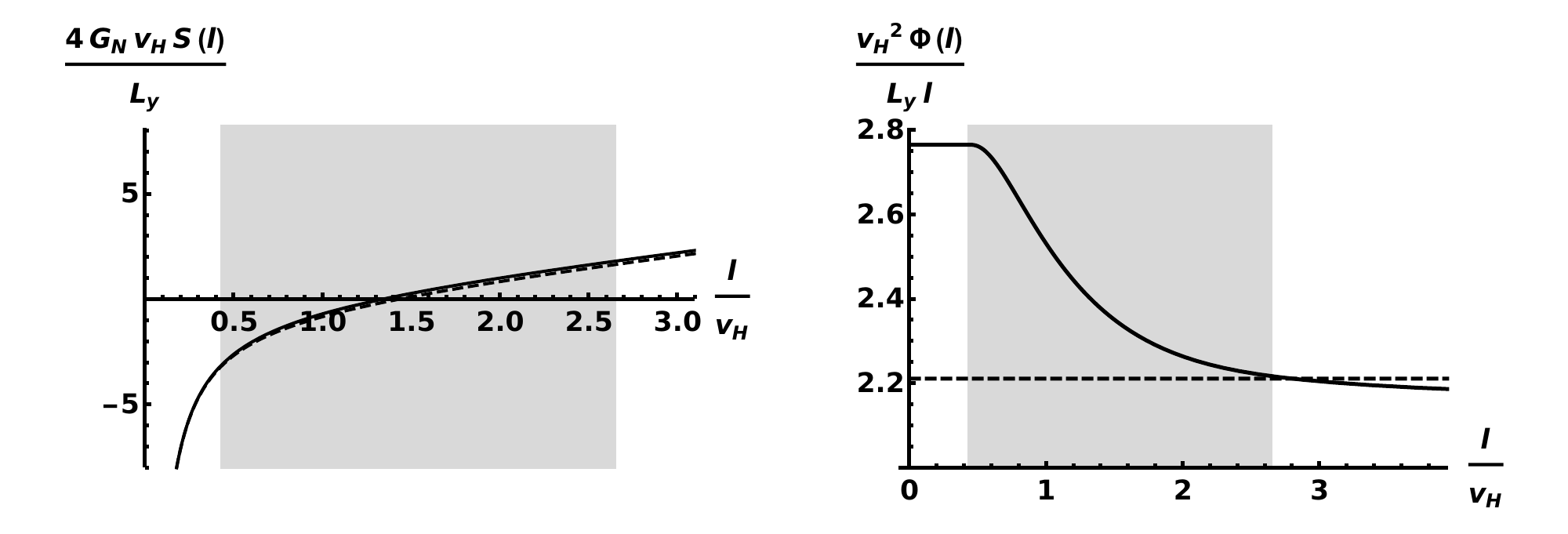}
  \caption{\small{\bf Left}: Entanglement entropy for the electron cloud geometry with $T/\mu=0.02$. The shaded gray area corresponds to the strip widths where the turning point is inside the cloud. In the UV, the entanglement entropy has the expected CFT behavior $S(l)\sim l^{-1}$, while in the IR, it scales as $S(l)\sim l$. The existence of the cloud can be inferred from the temperature dependence of the multiplicative constant in this latter regime, as is discussed in the text. {\bf Right}: Flux through the strip per unit width. Again the shaded gray area corresponds to the cloud region. It can be seen that when the strip lies outside the cloud (small $l/v_H$) the flux has the expected behavior $\Phi(l)\sim l$. The coefficient of proportionality is directly related to the total charge density of the black hole and cloud. In the IR region, the strip has dropped below the cloud, so the same extensive behavior is recovered, now with a coefficient that counts only the charge originating from the black hole. For the sake of comparison, in both figures we have also shown the results that are obtained in the pure RN case (dashed lines).}
  \label{fig:hee}
\end{figure}

With the above definitions in mind, we can now analyze the results for the entanglement entropy, which are presented in Fig.~\ref{fig:hee}. It behaves in the expected way in various regimes: In the UV $(lT_{\text{eff}}\ll1)$, the bulk geometry is AdS$_4$ and correspondingly, the entanglement entropy behaves as
\be
S(l) \approx -\frac{c_1}{l}\,,
\ee
with a coefficient $c_1>0$ that is independent of the temperature $T$ or chemical potential $\mu$. This holds for both, the electron cloud and AdS$_4$-RN solutions. In the IR $(lT_{\text{eff}}\gg1)$, thermal correlations dominate and the entropy becomes extensive, {\emph{i.e.}}, $S(l)\propto s_{\text{th}}$. In this case the RT surface tends to wrap part of the horizon and we expect that $S(l)\propto T_{\text{eff}}^2l$. More specifically, for the electron cloud we find that
\be\label{eecloud}
S(l)\approx \begin{cases}
  \displaystyle c_2 T^{2/z}\mu^{2-2/z} l  & \displaystyle \quad (T/\mu\ll1)\,,\\[2ex]
  \displaystyle c_3 T^2l & \displaystyle \quad (T/\mu\gg1)\,,
 \end{cases}
\ee
with $c_2>0$ and $c_3>0$. The temperature dependence at $T/\mu\gg1$ is trivial, since in this regime the cloud no longer exists and we recover the standard RN results. On the other hand, the temperature dependence at $T/\mu\ll1$ differs from the expected result in a pure RN black brane \cite{Kundu:2016dyk}, in which case one finds $S(l)\approx c_4 \mu^2l$. This difference hints (although indirectly) the existence of the cloud and a backreacted IR region. Finally, we point out that we could also diagnose the edges of the cloud by tracking down jumps in derivatives (of sufficiently high order) of the entanglement entropy with respect to the width $l$. This is, however, a feature of this particular model (because the cloud has exact compact support) but does not extend to more general fractionalized states. One might imagine, for instance, states dual to charged fluid distributions with tails that extend throughout the bulk. In these cases, then, one would find that continuity across scales is restored.\footnote{Discontinuities in derivatives also appear for mutual information, entanglement of purification, and generalized entanglement. The same phenomenon was discussed in the presence of a magnetic field in \cite{Puletti:2015gwa}. Here, we do not put much emphasis on this because these jumps can only be attributed to the particular model, and are not reminiscent of general fractionalized states.}

Before closing this section, let us offer some comments about the electric flux that goes through the RT surfaces discussed above. Using the known profile $\hat x'(\hat v)$, this flux can be written as
\begin{align}
  \frac{v_H \Phi(\hat v_*)}{L_y} = 2 \int_0^{\hat v_*} \frac{\hat v^3}{\hat v_*^2} \frac{\sqrt{\hat g(\hat v)} Q(\hat v)}{\sqrt{1-\frac{\hat v^4}{\hat v_*^4}}} d\hat v \ ,
\end{align}
where $Q(\hat v)$ is the integrated charge density below $v$. It can be computed from
\begin{align}
  Q(\hat v) &= Q_{BH} + \int_{\hat v}^1 \frac{\sqrt{\hat g(s)}}{s^2} \hat \sigma(s) ds \,,\nonumber\\
  &= Q_{BH} + \int_{\hat v}^1 \frac{d}{d s} \left( \frac{\hat h'(s)}{s^2 \sqrt{\hat f(s) \hat g(s)}} \right) ds \,,\nonumber\\
  &= - \frac{\hat h'(\hat v)}{\hat v^2 \sqrt{\hat f(\hat v) \hat g(\hat v)}} \ \label{eq:integrated-charge-density},
\end{align}
where on the second line we substituted equation \eqref{eq:sigma}. By the Gauss law, the electric flux must behave extensively in the strip width in both UV and IR limits. This happens in the UV because the strip is completely outside the cloud, and thus the flux is extensive in the strip width. In the IR, the strip minimal surface dives through the bulk and starts to trace the black hole horizon. In this limit, the flux counts only the charge originating from the horizon and is again extensive in strip width. This intuition is indeed confirmed by explicit calculations, as illustrated in Fig.~\ref{fig:hee}. The point to make here is that, even though the change in electric flux is substantial as we compare the cloud and RN solutions, the RT surfaces seem to be largely insensitive to it. Indeed, the RT surfaces only care about the bulk geometry, and not about the matter that is placed there, either charged or uncharged. Moreover, the geometry is only affected by the cloud through the effects of backreaction, which are highly suppressed. This observation is the main motivation for our proposed generalized entanglement functional, which we will discuss in section \ref{sec:generalized_functional}.

\subsection{Mutual information}

The mutual information is a correlation measure between two subsystems, $A$ and $B$, built out of entanglement entropies:
\begin{align}
  I(A,B) = S(A) + S(B) - S(AB) \ .
\end{align}
The individual entanglement entropies are computed with the same holographic formula as in the previous subsection. By holographic considerations, it is easy to see that the mutual information is UV finite by construction. It is also non-negative by subadditivity. The last term $S(AB)$, is in either the disconnected phase or the connected phase. In the disconnected phase $S(AB) = S(A) + S(B)$, and the mutual information vanishes, so $I(A,B)$ is non-zero only in the connected phase. These two possible phases are illustrated in left panel of Fig.~\ref{fig:ew}.

\begin{figure}[t!]
  \centering
  \includegraphics[width=1.0\textwidth]{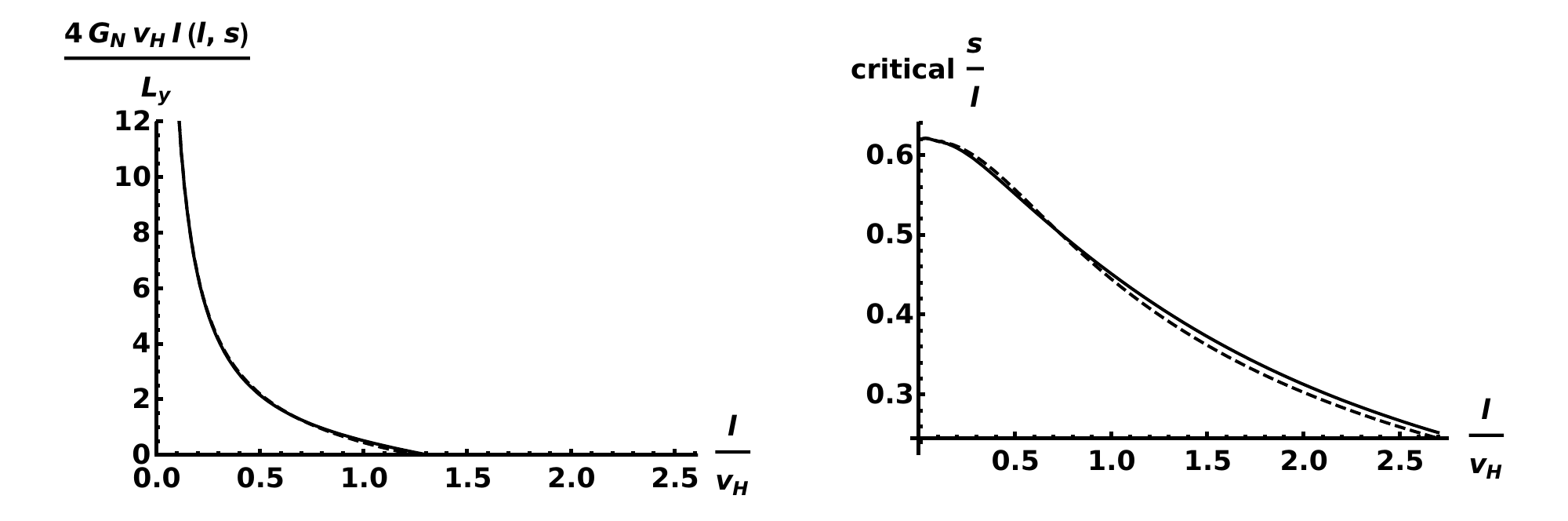}
  \caption{\small{\bf Left}: Mutual information for the electron cloud geometry with $T/\mu = 0.02$ corresponding to the parallel strip configuration with separation to length ratio $s/l=0.4$. The point where $I(A,B)$ vanishes, corresponds to the phase transition between connected and disconnected phases. {\bf Right}: The critical ratio $s/l$ at which the phase transition happens as a function of $l$. In the UV, the ratio approaches the CFT value $(\sqrt{5}-1)/2$. In the IR, the critical ratio tends to zero. Again, for comparison's sake, in both figures we have also shown the results that are obtained in the pure RN case (dashed lines).}
  \label{fig:mutual_info}
\end{figure}

The mutual information is straightforward to compute for parallel strips when we already know the strip entanglement entropy. This is because, by symmetry, we can express the entanglement entropy of any union of parallel strips $S(AB)$ as a sum of single strip entanglement entropies. Below, we will consider a symmetric case where the two strips have width $l$ and are separated by a distance $s$. Configurations of non-equal strips are also easy to work with, but we restrict ourselves to this symmetric configuration because in the next section, when we compute the entanglement wedge cross section, the expressions are greatly simplified in cases with this symmetry. Though, we note that a general algorithm for non-equal strips was worked out in \cite{Liu:2019qje}.

On general grounds, it is expected  that the mutual information is non-vanishing when the strip separation $s$ is small enough compared to their size $l$. For large separations on the other hand, we expect for the mutual information to vanish. This is exactly the behavior we find in Fig.~\ref{fig:mutual_info}. Here we have fixed the strip separation to $s/l=0.4$. The right panel of Fig.~\ref{fig:mutual_info} shows the critical separation $s/l$ where the connected/disconnected phase transition happens. Since our spacetime is asymptotically AdS$_4$, we expect for the critical $s/l$ to tend to the CFT value, which is given by the inverse golden ratio $(\sqrt{5}-1)/2$ \cite{Balasubramanian:2018qqx}. In the IR on the other hand, the critical $s/l$ should tend to zero, since in the planar black hole there exists a separation $s_\text{crit}$ such that for $s > s_\text{crit}$ no connected phase exists, even when $l\to\infty$\cite{Yang:2018gfq,Jokela:2019ebz}.

It is interesting to ask about how the mutual information can be used to diagnose the existence of the cloud. In order to do so, we need to be in an appropriate regime such that at least one of the RT surfaces that is used to compute $I(A,B)$ probes the deep IR geometry, yet the cloud solution still dominates over the RN solution. Furthermore, we also require that in such a regime the connected phase is the relevant one, so that the mutual information is non-vanishing. A careful inspection shows that the regime that we are interested in is when $T/\mu\ll1$, $lT_{\text{eff}}\gg1$ and $sT_{\text{eff}}\ll1$ (the latter two implying $l/s\gg1$). If this is satisfied, then, the scaling of $I(A,B)$ with respect to the temperature can be extracted from the leading UV and IR behavior of the entanglement entropies that enter the calculation. More specifically, we find that such a regime, the mutual information for the cloud geometry reads
\bea\label{MIscaling}
I(A,B)=2S(l)-S(2l+s)-S(s)&\approx& 2c_2 T^{2/z}\mu^{2-2/z}l-c_2T^{2/z}\mu^{2-2/z}(2l+s)+\frac{c_1}{s}\,,\nonumber\\
&\approx&-c_2T^{2/z}\mu^{2-2/z}s+\frac{c_1}{s}\,,
\eea
where $c_1$ and $c_2$ are (dimensionful) constants which are independent of $T$, $\mu$, $l$, and $s$. In contrast, if we repeat this exercise in the pure RN case we find that $I(A,B)\approx-c_4\mu^{2}s+\frac{c_1}{s}$. We note that in this regime the dependence on $l$ drops out in both cases. If we fix $\mu$ and $s$, and let $T$ vary, we can easily diagnose the existence of the cloud by tracking down the dependence of $I(A,B)$ with $T$. This is completely analogous to the analysis presented in the previous subsection based on entanglement entropies. As a final remark, we point out that also here one can expect that some appropriate derivatives of the mutual information (both with respect to $s$ and with respect to $l$) will jump discontinuously, as the relevant RT surfaces cross the edges of the cloud. This could help to diagnose the position of the cloud in the bulk; however, as explained in the previous section, we must emphasize that these jumps can only be attributed to the particular cloud model and are not to be associated with general fractionalized states.

\subsection{Entanglement of purification}
We now turn to the calculation of entanglement of purification. The proposed gravity dual for this quantity is given by the entanglement wedge cross section $E_W$ \cite{Takayanagi:2017knl,Nguyen:2017yqw}.\footnote{Note that there are many other CFT quantities that have also been linked to $E_W$, including logarithmic negativity \cite{Kudler-Flam:2018qjo}, odd entropy \cite{Tamaoka:2018ned}, entanglement distillation \cite{Agon:2018lwq} and reflected entropy \cite{Dutta:2019gen}. Among these, the last one is most often discussed in the literature, partly because its CFT counterpart is generally easier to compute. However, a challenge that remains to be addressed is the non-monotonicity of $E_W$ when conformal symmetry is broken \cite{Jokela:2019ebz}.}
The entanglement of purification is a correlation measure for mixed states. It is defined by
\begin{align}
  E_P(A,B) = \!\min_{\ket\Psi_{AA'BB'}}\! S(AA') \ ,
\end{align}
where the minimization is over all purifications $\ket\Psi$ of $\rho_{AB}$ and $S(AA')$ is the usual von Neumann entropy. For pure states, this reduces to the entanglement entropy $S(A)$.

\begin{figure}[t!]
  \centering
  \includegraphics[width=1.0\textwidth]{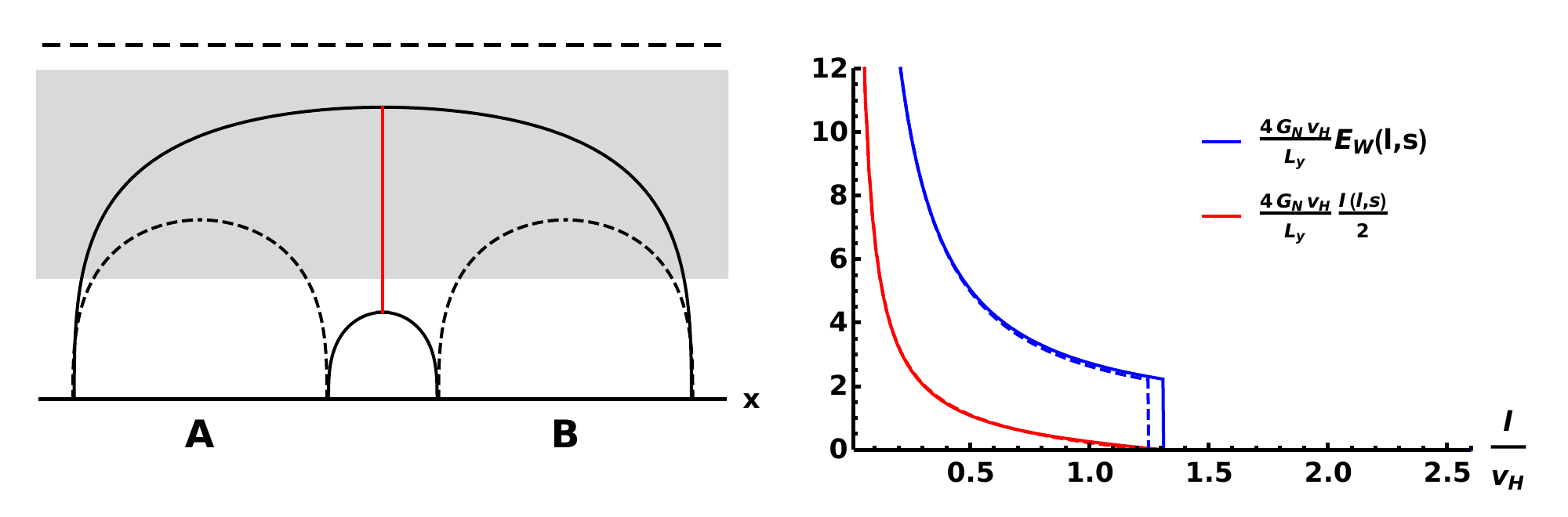}
  \caption{\small{\bf Left}: Minimal surfaces in the parallel strip case for the electron cloud geometry with $T/\mu = 0.02$. The disconnected phase of $S(AB)$ corresponds to the union of the dashed curves. The connected, $I(A,B)>0$, phase on the other hand corresponds to the solid black curves. The dashed black line on the top denotes the black hole horizon while the gray shaded area marks the location of the cloud. The red vertical line corresponds to the minimal surface whose area computes the entanglement wedge cross section in the connected phase. {\bf Right}: Mutual information and entanglement of purification for a parallel strip configuration with separation $s/l=0.4$, as a function of strip width $l$. The point where $I(A,B)$ vanishes corresponds to the disconnecting of the entanglement wedge. Thus at this point, mutual information vanishes continuously and the entanglement wedge cross section jumps to zero discontinuously. We also show the results for mutual information and entanglement of purification obtained in the pure RN case (dashed red and blue lines, respectively).}
  \label{fig:ew}
\end{figure}

For this calculation, we consider two disjoint regions $A$ and $B$ on the boundary. The information contained in the reduced density matrix $\rho_{AB}$ of the bipartite system is encoded in the entanglement wedge in the bulk. Since we are working in a static situation, the entanglement wedge is the bulk region bounded by the minimal surface associated with $S(AB)$. The entanglement wedge cross section $E_W$ is then the minimal area of a surface that splits the wedge into two parts, one part associated with $A$ and one part associated with $B$. If $S(AB)$ is in its disconnected phase, $E_W$ vanishes trivially, since the entanglement wedge separates automatically. In the connected phase of $S(AB)$ we are to scan over all possible splits of the entanglement wedge and select the one with minimal area\footnote{Alternatively, one can consider relaxing this latter  minimization, in which case the area of the bulk surface still gives an entanglement entropy in the optimal purification, but with a different bipartition of the purifying degrees of freedom \cite{Espindola:2018ozt}.}
\begin{align}
  E_P(A,B)= E_W(A,B) \equiv \min_\text{all splits} \frac{\text{Area}(\Gamma_{AB})}{4 G_N} \ ,
\end{align}
where $\Gamma_{AB}$ is the surface that splits the entanglement wedge. See Fig.~\ref{fig:ew} for an illustration of the entanglement wedge cross section.

In general, it is difficult to tell which surface $\Gamma_{AB}$ splits the entanglement wedge with minimal surface area. To overcome this problem, we study again the case of parallel, infinitely long strips with equal width. In this symmetric case, the minimal split is a surface that cuts the wedge at its symmetry axis, as shown in the left panel of Fig.~\ref{fig:ew}.

The entanglement wedge cross section for this configuration is then
\begin{align}\label{eq:EW}
  E_W(A,B) = \frac{L_y}{4 G_N v_H} \int_{\hat v_*^{(1)}}^{\hat v_*^{(2)}} \frac{\sqrt{\hat g(\hat v)}}{\hat v} d\hat v \ ,
\end{align}
where $\hat v_*^{(1)}$ and $\hat v_*^{(2)}$ are the turning points of strips of with $s$ and $2l+s$, respectively, solvable from \eqref{eq:strip_width}. This expression holds only when $I(A,B) > 0$, that is, we are in the connected phase of $S(AB)$. Otherwise, $E_W=0$. We have plotted $E_W$ in the right panel of Fig.~\ref{fig:ew} for strips of width $l$ and with the strip separation fixed to $s/l=0.4$. It can be seen that in the IR, correlation between strips vanish, as measured by $I(A,B)$ and $E_P(A,B)$. The figure also confirms the proved inequality
\begin{align}
  E_P(A,B) \geq \frac{1}{2} I(A,B)
\end{align}
obeyed by the entanglement wedge cross section.

Following the analysis of the previous two observables, we can also ask if the entanglement of purification can be used to diagnose the existence of the cloud. In this case, the situation is very similar to that of the mutual information, because the RT surfaces that define the entanglement wedge are the same to those that compute $I(A,B)$. It can be checked that the regime where $T/\mu\ll1$, $lT_{\text{eff}}\gg1$ and $sT_{\text{eff}}\ll1$ (implying $l/s\gg1$) is also well suited here: the entanglement wedge probes the deep IR of the geometry while being in its connected phase. Moreover, the cloud solution still dominates over the RN solution. The calculation of $E_P(A,B)$, however, involves different ingredients and cannot be computed solely from entanglement entropies. Fortunately, analytic expansions in various regimes of interest have been worked out in \cite{BabaeiVelni:2019pkw}. Here we will merely transcribe results that are relevant to us. In particular, for a theory with Lifshitz scaling, {\emph{i.e.}}, valid for the cloud in the regime where $T/\mu\ll1$) and $l/s\gg1$ we expect that:\footnote{Notice that we have changed the UV contribution, {\emph{i.e.}}, the last term in (\ref{ewcsres}), to account for the fact that the cloud is asymptotically AdS$_4$. In addition, we have included $\mu$-dependent factors in the IR terms that ensure that the constants $\tilde{c}_i$ share the same units. }
\be\label{ewcsres}
E_P(A,B)\approx \tilde{c}_1T^{1/z}\mu^{1-1/z}+\tilde{c}_2 s^{1+z}T^{1+2/z}\mu^{1+z-2/z}-\frac{\tilde{c}_3}{s}\,,
\ee
where $\tilde{c}_1$, $\tilde{c}_2$ and $\tilde{c}_3$ are (dimensionful) constants which are independent of $T$, $\mu$, $l$ and $s$.  In contrast, in standard RN case one expects that in this regime $E_P(A,B)\approx \bar{c}_1\mu+\bar{c}_2 s^{2}\mu^{3}-\frac{\tilde{c}_3}{s}$.\footnote{To find this result we have let $z\to1$ and $T\to T_{\text{eff}}\to\mu$. We also note that $\bar{c}_1\neq \tilde{c}_1$, $\bar{c}_2\neq \tilde{c}_2$.}  Again, we conclude that by carefully characterizing the temperature dependence of  $E_P(A,B)$ in this regime, we should be able to detect the subtle differences between the cloud and the RN solution, thus, recognizing the existence of the cloud. In addition, jumps in derivatives with respect to $s$ or $l$ could help diagnosing the position of the cloud in the bulk, at least for this particular model. More generally, this last feature will not show up in more general fractionalized states where the bulk charge is distributed everywhere in the bulk.

\subsection{Butterfly velocity}

Another interesting information theoretic observable that could yield further insights into the characterization of fractionalized phases is the so-called butterfly velocity $V_B$. This quantity is often discussed in the study of many-body quantum chaos and it is useful to diagnose how quickly the system responds after the insertion of local perturbations. Given a pair of generic Hermitian operators $W$ and $V$, the butterfly velocity is defined through the commutator \cite{larkin}
\be
C(t,\vec{x})=- \langle [W(t,\vec{x}),V(0,0)]^2 \rangle\,.
\label{eq-C(t,x)pre}
\ee
For quantum chaotic systems, this quantity is expected to grow as
\begin{equation}
C(t,\vec{x}) \sim \frac{1}{N^2} \exp \left[ \lambda_L\left(t-\frac{|\vec{x}|}{V_B}\right) \right]\,, \qquad |\vec{x}|\gg1/T\,,\quad t\gg1/T \,,
\label{eq-C(t,x)}
\end{equation}
The Lyapunov exponent $\lambda_L$ appearing in the exponential is a signature of \emph{fast scrambling}, and has been proven to have an upper bound for general quantum systems \cite{Maldacena:2015waa},
\be\label{boundchaos}
\lambda_L\leq 2\pi T\,.
\ee
Strikingly, this bound is sharply saturated for a number of systems, including strongly coupled field theories with Einstein gravity duals\footnote{Open-closed string duality in turn implies that bound is also saturated in the open string sector \cite{deBoer:2017xdk,Murata:2017rbp,Banerjee:2018twd}.} \cite{Shenker:2013pqa,Shenker:2014cwa} as well as ensemble theories
such as the Sachdev--Ye--Kitaev model and its cousins \cite{Maldacena:2016hyu,Gu:2016oyy,Davison:2016ngz}. The butterfly velocity $V_B$ characterizes the rate of expansion of $W$ due to a local perturbation caused by $V$. This quantity defines an \emph{emergent} light cone $\Delta t = |\vec{x}|/V_B$ such that within the cone $C(t,\vec{x}) \sim \mathcal{O}(1)$, whereas outside the cone $C(t,\vec{x}) \approx 0$. Based on this observation, \cite{Roberts:2016wdl} argued that, in holographic theories, $V_B$ acts as a low-energy Lieb-Robinson velocity $V_{LR}$ which limits the rate of transfer of quantum information. However, contrary to $\lambda_L$ (\ref{boundchaos}), there is no known universal bound for $V_B$ that holds generally \cite{Giataganas:2017koz,Fischler:2018kwt,Gursoy:2020kjd}. On the other hand, there is an interesting relation between the butterfly velocity and transport, that can be derived from universal properties of black hole horizons \cite{Blake:2016wvh,Davison:2016ngz,Blake:2017qgd}. On general grounds, one expect that charge and energy diffusion constants to scale as
\be\label{eq:diffD}
D_c=C\frac{\hbar V^2}{k_B T}\,,\qquad D_e=E\frac{\hbar V^2}{k_B T}\,,
\ee
where $C$ and $E$ are constants and $V$ is a characteristic velocity of the theory. In \cite{Blake:2016wvh,Davison:2016ngz,Blake:2017qgd} it was argued that a natural candidate for such a velocity in a theory without quasi-particle excitations is provided by the butterfly velocity $V_B$. For theories with a particle-hole symmetry, both relations were proved true, with $C$ and $E$ taking universal values that depend on the universality class of the theories in the IR. However, for finite density states the energy and charge currents overlap and, as a result, only the latter statement remains true in general.\footnote{Some finite density systems still satisfy the left equation in (\ref{eq:diffD}) suggesting an approximate particle-hole symmetry in the IR \cite{Blake:2016sud}.} In this case, one  would  expect  to  move  away  from  the  universal  regime and
(\ref{eq:diffD}) would only provide a \emph{bound on charge diffusion} \cite{Hartnoll:2014lpa}. Nevertheless, we will show below that at least in the zero temperature limit, the above connection will allow us to infer the existence of bulk charges and distinguish them from those hidden behind black hole horizons.

Following the ideas of \cite{Mezei:2016wfz}, it can be shown that the butterfly velocity $V_B$ can be derived using simple ideas of subregion duality and entanglement dynamics. More specifically, the method for deriving $V_B$ proposed in \cite{Mezei:2016wfz} amounts to add a local perturbation in the CFT and then follow the time-like trajectory it traces out in the bulk with entanglement surfaces. At late times, the perturbation is red-shifted from the point of view of an observer at the AdS boundary, and the entangling surfaces start to sweep the black hole horizon, leading to longer and longer regions in the CFT. The rate at which these regions increase give the butterfly velocity, which turns out to determined in terms of near-horizon data only. For $4$-dimensional bulk geometries it reduces to (see Appendix \ref{app:vb} for details)
\be\label{eq:butt}
V_B=\sqrt{\frac{g'_{tt}(v_H)}{2g_{ii}'(v_H)}}=\sqrt{\frac{\hat{g}'_{\hat{t}\hat{t}}(1)}{2\hat{g}_{\hat{i}\hat{i}}'(1)}} \ .
\ee

\begin{figure}[t!]
  \centering
  \includegraphics[width=1.0\textwidth]{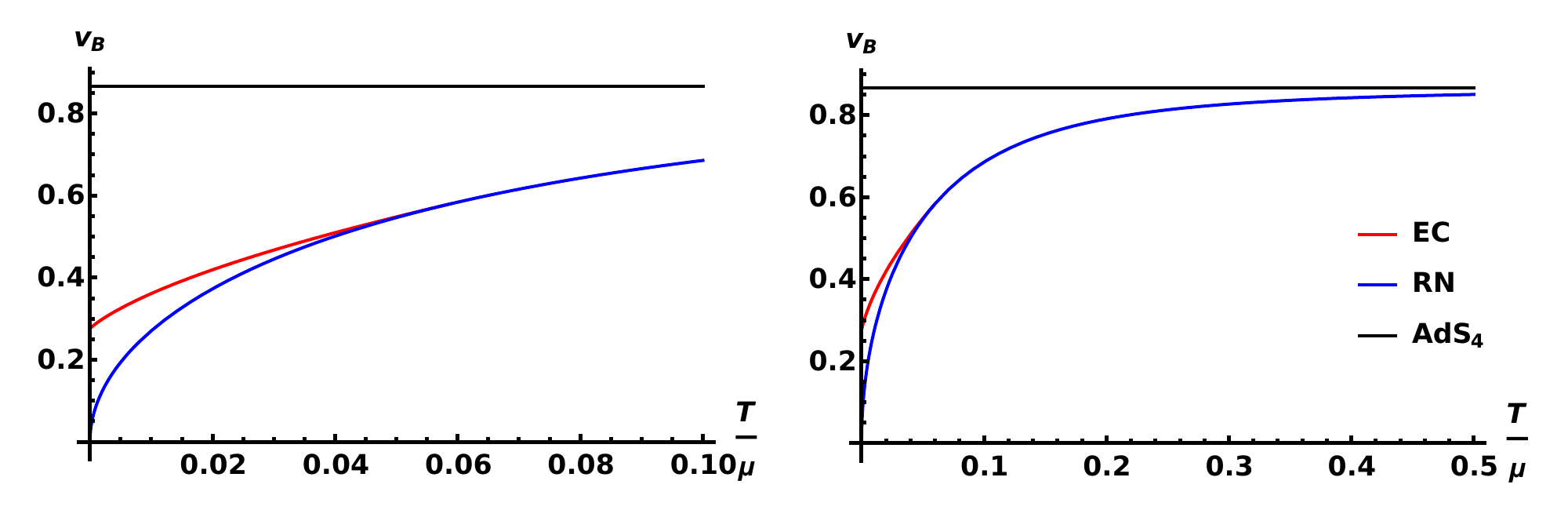}
  \caption{\small{\bf Left:} We depict the butterfly velocity (\ref{eq:butt}) as a function of $T/\mu$. Notice that at high temperature we obtain the black brane result $\sqrt{3/4}$ shown as black line, while below the phase transition the velocity is always higher than that for a AdS-RN background. {\bf Right:} Same quantity as on the left panel, but shown for larger values of $T/\mu$.}
  \label{fig:butterfly}
\end{figure}
In Fig.~\ref{fig:butterfly} we have depicted the butterfly velocity (\ref{eq:butt}) in the cloud geometry. We note that the velocity asymptotes to $V_B\to\sqrt{3/4}\approx 0.866$ at high temperature, as expected. This value can be derived from fact that conformal invariance is restored in the UV. Interestingly, in the opposite limit, the butterfly velocity saturates to a non-zero value, in stark contrast to pure RN case. Given that the butterfly velocity provides a bound on charge diffusion \cite{Blake:2016wvh,Davison:2016ngz,Blake:2017qgd,Hartnoll:2014lpa}, this non-zero value can then be attributed to the existence of dissipationless charged degrees of freedom hovering outside the horizon, {\emph{i.e.}}, cohesive charges. More specifically, one can show that the butterfly velocity can be written in terms of the charge parameter of the inner solution $\hat{q}$ as
\be\label{VBgeneral}
V_B=\sqrt{\frac{3}{4}\left(1-\frac{\hat{q}^2}{6}\right)}\,.
\ee
In the absence of a cloud, one find that $V_B\to0$ as $T/\mu\to0$ ($\hat{q}\to\sqrt{6}$), {\emph{i.e.}}, as the black hole becomes extremal. This signalizes the dissipative nature of the active degrees of freedom in the IR (note that in this case the bound on charge diffusion is exact: $D_c=0$). However, in the presence of a cloud, $\hat{q}$ is not directly related to the physical charge, because it is screened by the cloud. Instead, one finds that $\hat{q}\to \hat{q}_0<\sqrt{6}$, as we let $T/\mu\to0$, so $V_B$ remains finite. This in turn indicates the existence of cohesive charges in the bulk. Finally, we point out that since the phase transition between the cloud and the RN solution is of third order, we expect that
\be\label{vBtransition}
\Delta V_B=\left(V_B\right)_{\text{AdS--RN}}-\left(V_B\right)_{\text{EC}}\propto(T_c-T)^2\,.
\ee
The exponent in (\ref{vBtransition}) is due to the fact that the computation of $V_B$ requires first order derivatives of the metric functions, hence it is lowered by one. Indeed, we can confirm the above scaling from our numerical results.

\section{Generalized entanglement functional: a refined diagnostic of fractionalization\label{sec:generalized_functional}}

\subsection{Coarse grained entanglement entropy}

In the previous section we studied codimension-2 bulk surfaces whose areas give entanglement entropies of boundary regions. We also computed the electric flux through these surfaces, and showed that it has a negligible effect on the shape and area of the surfaces. In this section we will study codimension-2 surfaces governed by a more general functional which do take into account the explicit effects of the electric flux. The specific choice of functional is motivated in Appendix~\ref{sec:functional_derivation} and follows from the application of the Iyer--Wald formalism for a theory of gravity coupled to a $U(1)$ gauge field. The calculation is rather technical, so for the sake of simplicity we will merely state the final result here.  The general functional that we obtain is given by (\ref{eq:functionalgen}), \emph{i.e.},
\be\label{eq:functionalgen}
\mathcal{S}(A)=-2\pi \int_{\tilde{\Gamma}_A}\left(\frac{\partial {\cal L}}{\partial R_{\mu\nu\alpha\beta}}\mbox{{\boldmath
$\epsilon$}}_{\mu\nu}\mbox{{\boldmath
$\epsilon$}}_{\alpha\beta}-\gamma\frac{\partial {\cal L}}{\partial F_{\alpha\beta}}\mbox{{\boldmath
$\epsilon$}}_{\alpha\beta} \right)\ ,
\ee
where ${\cal L}$ is the combined Lagrangian for gravity and matter fields, $\gamma\equiv \xi\cdot A$ and $\xi$ is a bulk Killing vector, which for static spacetimes can be taken to be $\xi=\partial_t$. Specializing to the case of pure electric fields, we find a simplified version, equation (\ref{eq:functional}), which can be written schematically as the sum of area and flux terms studied in the previous section
\begin{align}\label{eq:functionalM}
  \mathcal{S}(A) = \frac{1}{4 G_N} \text{Area}(\tilde{\Gamma}_A)+ \tilde{\Phi}(\tilde{\Gamma}_A) \ ,\qquad \tilde{\Phi}(\tilde{\Gamma}_A)\equiv-\frac{2\pi}{e^2}\int_{\tilde{\Gamma}_A}\gamma\, E_\perp \ .
\end{align}
Importantly, $\tilde{\Phi}$ here is a \emph{normalized} flux with a weighting factor given by the local chemical potential (\ref{eq:muloc}), which in our setup is given by
\begin{align}\label{eq:gamma}
  \gamma(v)  = \xi \cdot A = \frac{e}{\kappa}\frac{h(v)}{\sqrt{ f(v)}} = \mu_{loc}(v)  \ .
\end{align}
It is easy to see that (\ref{eq:functionalM}) reduces in the IR to the Hartnoll-Radi\v{c}evi\'c functional, proposed originally in \cite{Hartnoll:2012ux} as an order parameter for charge fractionalization. The only difference between the two prescriptions is that in their proposal $\gamma$ is taken to be a constant, so the bulk charges contribute equally regardless of their radial position in the bulk.

Let us offer a couple of comments about the generalized functional (\ref{eq:functionalM}). We demand that in the absence of any charges, the generalized functional reduces to the entanglement entropy. Per continuity, we will therefore also assume that the ``generalized'' minimal surfaces satisfy the homology constraint. In the context of black hole thermodynamics, the functional is meant to be evaluated at the bifurcate horizon $v=v_H$. Since $\xi$ vanishes there, the flux term cancels out and one ends up with the standard Wald term for black hole entropy. However, in the context of entanglement entropy, we actually need to evaluate the functional at a different bulk surface $\tilde{\Gamma}_A$, and hence the flux term can give a non-zero contribution. Here $\tilde{\Gamma}_A$ is a new codimension-2 bulk surface, anchored also on the boundary of the entangling region $\partial \tilde{\Gamma}_A = \partial A$ and homologous to $A$, but resulting from the minimization of the new functional (\ref{eq:functionalM}). As we show below, the main effect of the flux term is to repel the surface towards the boundary when compared to the corresponding RT surface, giving rise to a \emph{shadow} region in the deep IR. Intuitively, this happens because we are tracing over (part) of the fractionalized and coherent charge degrees of freedom as we increase the size of the region (in addition to tracing over all degrees of freedom of the complementary region $A^c$), giving rise to a coarser measure of entanglement for the subsystem.

As in the previous section, we focus on strip geometries, with $-l/2\leq x \leq l/2, -L_y/2\leq y \leq L_y/2$, because this case is computationally simple and exhibits all the novel features we want to showcase. Specifically, for this setup, our functional (\ref{eq:functionalM}) reduces to
\begin{align}
  \frac{4 G_N v_H \mathcal{S}}{L_y} = 2 \int_0^{\hat v_*} \left( \frac{ \sqrt{\hat g(\hat v)\hat v^2+\hat x'(\hat v)^2}}{\hat v^2} + \hat\gamma(\hat v) Q(\hat v) \hat x'(\hat v) \right) d\hat v \ , \label{eq:generalized_strip_entropy}
\end{align}
with $Q(\hat v)$ defined as in (\ref{eq:integrated-charge-density}). The rescaled field $\hat\gamma(\hat v)$ here is defined such that the relative factor between the area and flux terms is absorbed into the definition, $\hat \gamma (\hat v)= (\kappa/e)\gamma (\hat v)$.  Moreover, we have chosen to integrate over the branch where $\hat x'(\hat v) \leq 0$ and $0 \leq \hat x(\hat v) \leq \frac{l}{2v_H}$. We note that the flux term is UV-finite because the area term forces the minimal surfaces to have $\hat x'(\hat v)=0$ near the boundary. Hence, the only UV-divergences originate from the area term. We can isolate the divergence piece out as we did for the entanglement entropy \eqref{eq:strip_entropy},
\begin{align}
  \frac{4 G_N v_H \mathcal{S}}{L_y} = \frac{2}{\hat\epsilon} - \frac{2}{\hat v_*} + 2 \int_0^{\hat v_*} \left( \frac{ \sqrt{\hat g(\hat v)\hat v^2+\hat x'(\hat v)^2}}{\hat v^2} - \frac{1}{\hat v^2} + \hat\gamma(\hat v) Q(\hat v) \hat x'(\hat v) \right) d\hat v \ , \label{eq:generalized_strip_entropy_2}
\end{align}
where $\hat\epsilon$ is an UV-cutoff in the radial coordinate $\hat v$. As in the case of the entanglement entropy, we will consider the regularized version of this functional which we define by subtracting the divergent $2/\hat\epsilon$-term.

The profile of $\hat x'(\hat v)$ is determined by the minimization of the area and flux terms combined. As before, this functional does not depend explicitly on $\hat x(\hat v)$ so we have a conserved quantity,
\begin{align}
  \frac{\hat x'(\hat v)}{\hat v^2 \sqrt{\hat v^2 \hat g(\hat v)+\hat x'(\hat v)^2}} + \hat \gamma(\hat v) Q(\hat v) = -\frac{1}{\hat v_*^2} + \hat \gamma(\hat v_*) Q(\hat v_*) \ , \label{eq:turning_point}
\end{align}
where $\hat v_*$ denotes the turning point. The above equation can be solved for $\hat x'(\hat v)$ which gives the strip width as the following integral
\begin{align}
  \frac{l}{v_H} = 2\int_0^{\hat v_*} \frac{\hat v^3\sqrt{g(\hat v)}\mathcal{Q}(\hat v, \hat v_*)}{\sqrt{\hat v_*^4-\hat v^4 \mathcal{Q}(\hat v, \hat v_*)^2}} d\hat v \ . \label{eq:generalized_strip_width}
\end{align}
where we have defined:
\begin{align*}
  \mathcal{Q}(\hat v,\hat v_*) \equiv 1+v_*^2\left(\hat \gamma(\hat v)Q(\hat v)-\hat\gamma(\hat v_*)Q(\hat v_*) \right) \ .
\end{align*}
Similarly, plugging $\hat x'(\hat v)$ back into our functional we obtain an alternative expression for the generalized entanglement entropy, which is independent of the profile $\hat x(\hat v)$:
\begin{align}
  \frac{4 G_N v_H \mathcal{S}}{L_y} = \frac{2}{\hat \epsilon}-\frac{2}{\hat v_*}+2\int_0^{\hat v_*}\left(
    \frac{\sqrt{g(\hat v)}}{\hat v}\left(\frac{\hat v_*^2 - \hat\gamma(\hat v)Q(\hat v)\mathcal{Q}(\hat v, \hat v_*)\hat v^4}{\sqrt{\hat v_*^4-\hat v^4 \mathcal{Q}(\hat v, \hat v_*)^2}}\right)
    -\frac{1}{\hat v^2}
    \right) d\hat v \ . \label{eq:generalized_simplified}
\end{align}

\begin{figure}[t!]
  \centering
  \includegraphics[width=1.0\textwidth]{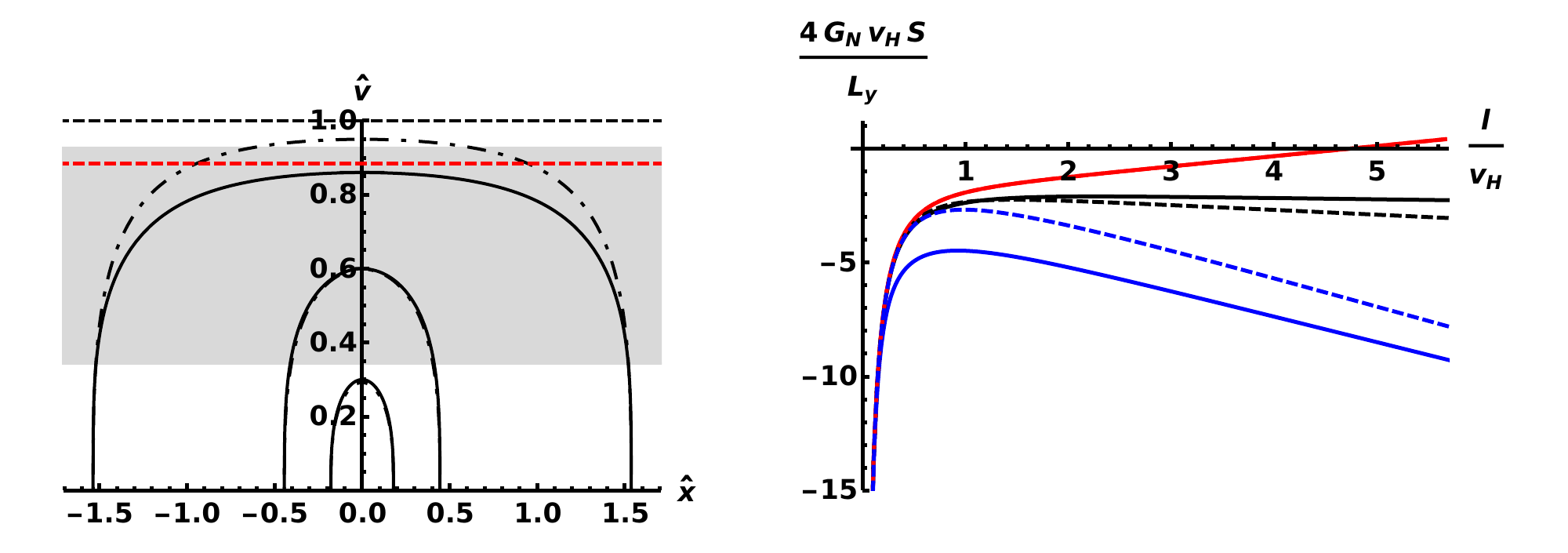}
  \caption{\small{\bf Left:} Minimal surfaces for a few boundary strips. The solid curves correspond to surfaces which minimize the generalized functional \eqref{eq:generalized_strip_entropy_2} while the dash-dotted curves represent RT surfaces of the same boundary strips. It can be seen that the generalized functional is repelled towards the boundary when compared to the corresponding RT surface, but this effect is visible only when the surface penetrates substantially into the cloud (shaded region). The horizontal dashed lines represent the black hole horizon (black) and the edge of the shadow region (red). For this figure we have set $T/\mu=0.02$. {\bf Right:} The value of the (regularized) generalized functional as a function of strip width. The different curves correspond to $T/\mu=0.0005$ (blue), $T/\mu=0.02$ (black), and $T/\mu=0.057$ (red). We see that strips with small lengths behave similarly for different $T/\mu$ but the long range slope is affected according to the formula \eqref{eq:asymptotic_slope}. For comparison, we also show the results for the generalized functional with the same values of $T/\mu$ in the pure RN case (dashed lines).}
  \label{fig:embeddings}
\end{figure}

\begin{figure}[t!]
  \centering
  \includegraphics[width=1.0\textwidth]{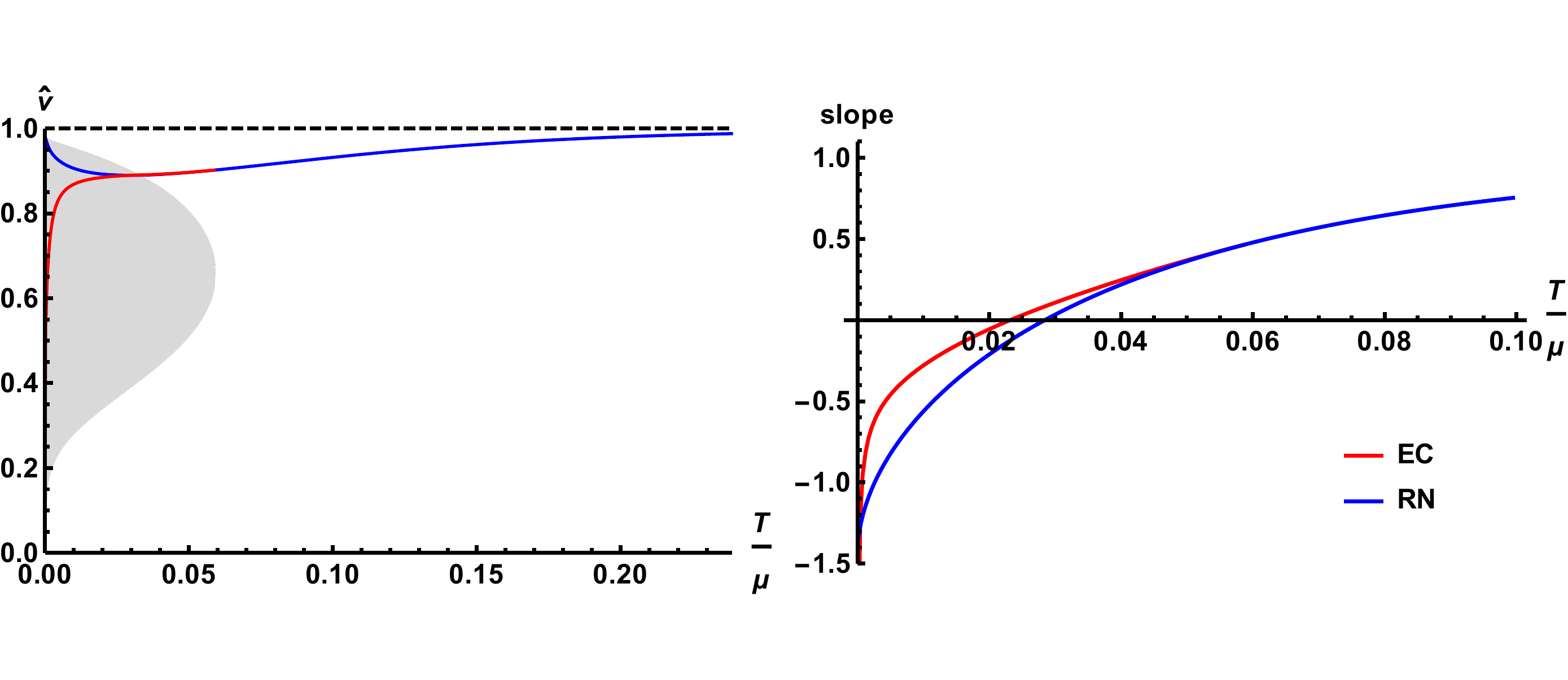}
  \caption{\small{\bf Left:} The curves indicate where the shadow region starts in our electron cloud background (red) and in AdS$_4$-Reissner--Nordstr\"om background (blue). The black dashed line indicates the black hole horizon. The gray area shows the size and position of the bulk charges of the electron cloud. {\bf Right:} The asymptotic slope of the generalized functional in the limit of large strip width \eqref{eq:asymptotic_slope}. In the Reissner--Nordstr\"om case, the slope approaches unity at high temperatures because the shadow approaches the event horizon where the weighting function $\hat\gamma(\hat v)$ vanishes leaving only the area contribution in \eqref{eq:asymptotic_slope}.}
  \label{fig:shadow_and_slope}
\end{figure}

A few examples of minimal surfaces that arise from our functional are shown in Fig.~\ref{fig:embeddings}, together with a plot of $\mathcal{S}(l)$. The first exceptional feature of the new functional can already be seen from the plots: the existence of a \emph{shadow}, {\emph{i.e.}}, a region in the bulk that cannot be probed by the generalized surfaces. This can be deduced from the above integrals (\ref{eq:generalized_strip_width})-(\ref{eq:generalized_simplified}). For instance, analyzing the square root in the denominator of (\ref{eq:generalized_strip_width}) we can determine value of $\hat v_*$ for which the integral diverges. We denote this value $\hat v_s$. It turns out that the profiles $\hat{x}(\hat v)$ are real only when $\hat v_* \leq \hat v_s$, with $\hat v_s<1$. In order to see this, consider setting $\hat v = \hat v_* - \lambda$ and then expand the argument of the square root in the denominator of (\ref{eq:generalized_strip_width}) for small $\lambda$,
\begin{align}
  0 + \lambda \frac{d}{d\hat v_*} \left( \hat \gamma(\hat v_*) Q(\hat v_*) - \frac{1}{\hat v_*^2} \right) + \mathcal O(\lambda^2) \ .
\end{align}
The vanishing of the zeroth order term follows from the definition of the turning point $\hat v_*$ \eqref{eq:turning_point}. The linear term is such that when moving from the boundary towards the horizon it is first positive and at some point it turns negative. However, this term cannot be negative because the integral \eqref{eq:generalized_strip_width} would become complex when integrating from the turning point $\hat v_*$ toward the boundary. Thus, the maximal value that $\hat v_*$ can attain occurs where the first order term vanishes. At this point $\hat v_*=\hat v_s$ and it corresponds to an infinitely wide surface, $l\to\infty$, hovering at a constant $\hat v=\hat v_s$. We can find the value of $\hat v_s$ as the root of
\begin{align}
  \frac{d}{d\hat v} \left( \hat \gamma(\hat v) Q(\hat v) - \frac{1}{\hat v^2} \right) = 0 \ . \label{eq:shadow_condition}
\end{align}
We call the bulk region $\hat v_s < \hat v < 1$ a \textit{shadow}, in analogy to the entanglement shadows which occur, {\emph{e.g.}}, for spherical black holes in AdS \cite{Freivogel:2014lja}.
We emphasize that the presence of a shadow is not a special feature of the bulk charges, but can be attributed to the coarse graining of the generalized entanglement. To see this, notice that the generalized functional experiences a shadow also in the Reissner-Nordstr\"om case where all charge is hidden behind the event horizon. Moreover, even though we found the shadow by studying strips, our numerics for disks suggest that the \emph{same} shadow is present also for other (sufficiently large) entangling regions, and thus is a property of the background (see Appendix \ref{app:disk} for details). The size and location of this shadow is shown in the left panel of Fig.~\ref{fig:shadow_and_slope}.

Another property of $\mathcal{S}(l)$ that is already visible from the plots is that for wide strips $l\gg v_H$, the generalized functional becomes linear in $l$ and, hence, extensive. Interestingly, the slope characterizing its IR behavior can have either sign. To understand this point we note that in this limit the value of the regularized functional can found by evaluating it on the bulk surface $\hat v=\hat v_s$. From \eqref{eq:generalized_simplified} it follows that
\begin{align}
  \frac{4 G_N v_H^2 \mathcal{S}}{L_y}\Bigg|_{l\to\infty} = \left( \frac{1}{\hat v_s^2} - \hat\gamma(\hat v_s) Q(\hat v_s) \right) l \ . \label{eq:asymptotic_slope}
\end{align}
The terms inside the parenthesis depend on the value of $T/\mu$, as illustrated in the right panel of Fig.~\ref{fig:shadow_and_slope}. Whether the regularized functional is a monotonous increasing function of $l/v_H$ or not depends on the relative importance of the area and flux terms. At large enough $T/\mu$ we find that $d\mathcal S/dl>0$ for all $l$. At low values of $T/\mu$, however, the flux contribution becomes more important and $\mathcal S$ starts to decrease for large strip widths.

\subsection{A $c$-function for cohesive charges\label{sec:c-func}}

Let us now discuss what could be a boundary measurement that could provide us with means to answer the question on the nature of charge carriers, whether they are subject to dissipation or not. To do so, we need to devise a probe that would distinguish between fractionalized charges and cohesive ones. The former are in one-to-one correspondence with charges behind the horizon in the dual gravity description, while the latter correspond to charges hovering above the horizon, {\emph{i.e.}}, those populating the cloud.

A natural way of addressing this type of questions is to construct a function that counts the number of degrees of freedom at different energy scales in the dual theory. For instance, a candidate for an ``entropic'' $c$-function that counts the \emph{total} number of degrees of freedom in a $(2+1)$-dimensional homogeneous and isotropic system is $c \propto l^2 d S(l)/d l$ \cite{Casini:2004bw,Nishioka:2006gr,Myers:2010tj,Myers:2012ed,Casini:2012ei,Liu:2012eea,Liu:2013una}, where $S(l)$ is the entanglement entropy for a strip of length $l$. This proposal has been tested in holographic duals of $(2+1)$-dimensional ABJM Chern-Simons field theories and shown to meet expectations \cite{Bea:2013jxa,Kim:2014qpa}: it is a monotonically decreasing function under RG flow and precisely matches the number of degrees of freedom at the fixed points from the matrix model field theory calculation \cite{Santamaria:2010dm}. Moreover, it has an obvious advantage over the entanglement entropy because it does not depend on the details of the UV regulator.

Since the calculation of entanglement entropy in AdS/CFT is remarkably simple, the study of holographic $c$-functions based on entanglement entropy have become increasingly popular in recent years. Akin to the entanglement entropy, holographic $c$-functions based on entanglement can directly probe the finite correlation length in the underlying quantum field theory \cite{Jokela:2020wgs} and, hence, reveal aspects of its phase diagram \cite{Klebanov:2007ws}. In addition to this, entropic $c$-functions can quantitatively expose conformal fixed points at intermediate energy scales \cite{Jokela:2020wgs} (perhaps even those lurking in the complex plane \cite{Gorbenko:2018ncu,Faedo:2019nxw}) and give complementary information on the underlying mechanism for the phase transitions \cite{Baggioli:2020cld,Ecker:2020gnw}. We point out that there are various proposals for extending holographic $c$-functions to anisotropic systems \cite{Cremonini:2013ipa,Chu:2019uoh,Ghasemi:2019xrl,Hoyos:2020zeg,Cremonini:2020rdx}, with potential applications in, {\emph{e.g.}}, heavy-ion collisions \cite{Arefeva:2020uec}. However, the lack of underlying Lorentz invariance in these setups means there is no general theorem to guarantee monotonicity \cite{Hoyos:2020zeg}.

Coming back to our problem, let us henceforth use the above discussion as an inspiration and define the following function\footnote{The quantities in the denominator are meant to be computed on a
\emph{reference} background with only fractionalized charges, {\emph{i.e.}}, a pure AdS--RN solution. The two backgrounds must have the same asymptotic values for all bulk fields, in particular, the same value for $A_t(v\to0)\to\mu$.}
\be\label{eq:cfunction}
 {\cal{C}} \equiv \frac{\mathcal{S}'(l)-S'(l)}{\mathcal{S}_{RN}'(l)-S_{RN}'(l)}\,,
\ee
where $S$ is the standard entanglement entropy (\ref{eq:strip_entropy}) and $\mathcal{S}$ is its generalized version (\ref{eq:generalized_strip_entropy}). This function is constructed having in mind the following properties:
\begin{itemize}
 \item It should be constant in the absence of cohesive charges.
 \item It should approach finite values in the UV and IR, $\mathcal{C}_{UV}$ and $\mathcal{C}_{IR}$, with $\mathcal{C}_{UV}\geq \mathcal{C}_{IR}$.
 \item It should be monotonically decreasing along the RG flow.
\end{itemize}
Let us discuss these points in more detail and explain the reasoning behind this proposal. First off, we want to pick up the contribution from the bulk charges only, so it is natural to consider the difference $l^2(\mathcal{S}'(l)-S'(l))$ to subtract the area term, at least approximately. However, this combination is problematic because $i)$ it depends non-trivially on $l$ even when there are no cohesive charges in the bulk, and $ii)$ it diverges in the IR since both surfaces tend to sweep black hole horizon leading to linear-in-$l$ dependence for $S(l)$ and $\mathcal{S}(l)$ (with different coefficients). To fix these issues, we thus include the terms in the denominator of (\ref{eq:cfunction}), which suffices to guarantee the first and second properties discussed above. Notice that in the absence of a cloud, the ratio in (\ref{eq:cfunction}) then evaluates to ${\cal{C}}=1$, which is a desired property in the case where all the charge reside behind the horizon. In addition, the ratio cancels out the $l$ factors in the IR, leading to a finite value for $\mathcal{C}_{IR}$. Finally, it can be shown that in the presence of bulk charges, (\ref{eq:cfunction}) decreases monotonically as a function of $l$ \emph{even in the regime where the minimal surfaces do not probe the cloud region in the bulk}.\footnote{Assuming that the difference in area terms is \emph{negligible}, the monotonicity should follow from Gauss's law in combination with the nesting property for the generalized surfaces, {\emph{i.e.}}, the increase in the size of the bulk region enclosed as we increase $l$. Here, we assume that the bulk charges have all definite sign (equal to the charge behind the horizon). For more general states with bulk charges of varying sign, the function (\ref{eq:cfunction}) does not need to be monotonic in the size of the region. These states would, nevertheless, suffer from obvious electric instabilities.} The reason for this is that for the electron cloud solution the exterior geometry is that of a RN black brane with the same chemical potential $\mu$ but with different charge parameters, $Q_{\text{EC}}\neq Q_{\text{RN}}$. Hence, the results for both $S(l)$ and $\mathcal{S}(l)$ in the cloud deviate from those in the RN black brane even for $l\ll v_H$ (the regime where the minimal surfaces do not reach the cloud). As a result, the $\mathcal{C}$-function encodes information about the cohesive charges even in the deep UV regime!
\begin{figure}[t!]
  \centering
  \includegraphics[width=0.48\textwidth]{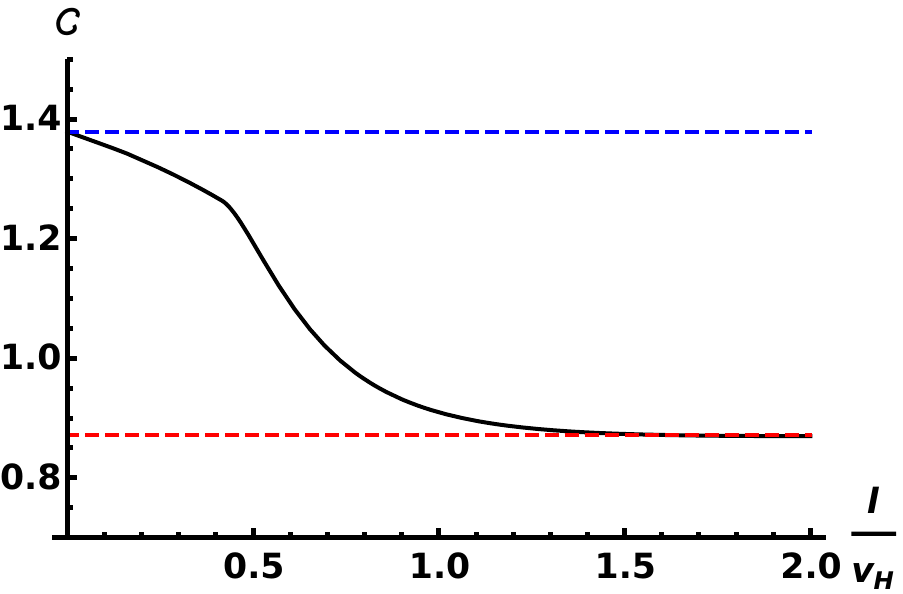}
  \includegraphics[width=0.48\textwidth]{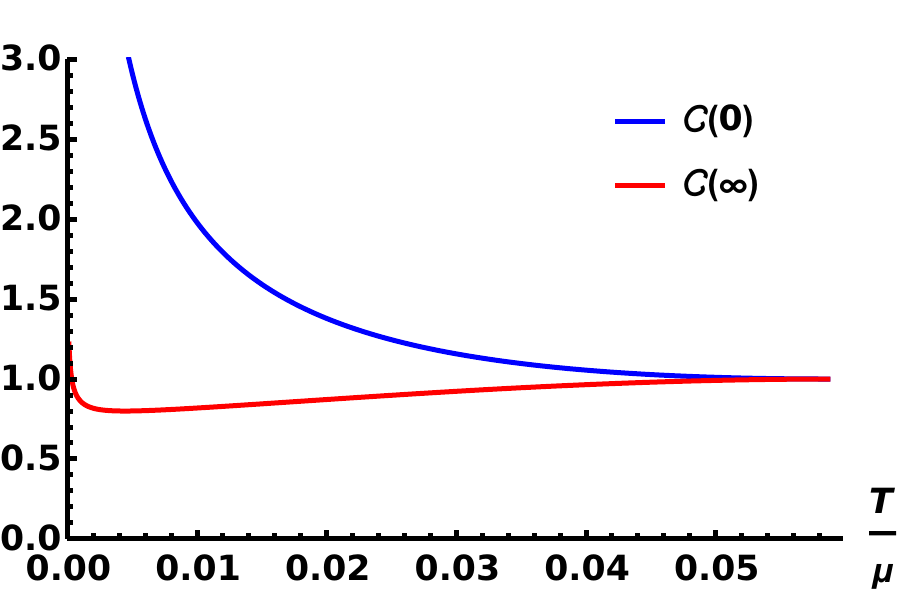}
  \caption{\small{\bf{Left:}} We depict the candidate $c$-function as defined in (\ref{eq:cfunction}) for fixed $T/\mu=0.02$. As expected, it will faithfully count the number of coherent charged degrees of freedom and will therefore only decrease in the scales when the minimal surfaces dive through the cloud region in the bulk. The blue/red dashed lines denote the UV/IR-limits of the proposed $c$-function which can be computed analytically. The slight knee visible in the curve around $l/v_H \approx 0.4$ corresponds to the point where the minimal surfaces start probing the cloud. {\bf{Right:}} We depict the effective number of degrees of freedom at the UV $\mathcal{C}_{UV}={\cal{C}}(0)$ and at the IR $\mathcal{C}_{IR}={\cal{C}}(\infty)$ as a function of $T/\mu$. Interestingly, $\mathcal{C}_{UV}$ can be shown to be proportional to the ratio between the total charge and the black hole charge (\ref{CUVpropto}), so it can be used as a probe to diagnose and quantify the existence of cohesive charges in the bulk.}\label{fig:cfunction}
\end{figure}

In Fig.~\ref{fig:cfunction} we have depicted the quantity (\ref{eq:cfunction}) for a small value of $T/\mu$, such that the electron star dominates over the RN solution. We find that this quantity indeed behaves as expected: it is monotonically decreasing as $l/v_H$ is increased, and approach to constant values both in the UV and IR. We also point out that $\mathcal{C}'(l)$ has a kink at exactly the value of $l$ at which the generalized surfaces start probing the cloud. However, this is a feature of this particular model (due to the compactness of the cloud sources) and will be softened in situations where the bulk charge is distributed smoothly across the bulk. Finally, in Appendix \ref{app:cfunctions} we derive analytic expressions for $\mathcal{C}_{UV}={\cal{C}}(0)$ and $\mathcal{C}_{IR}={\cal{C}}(\infty)$ in terms of various bulk parameters.
To do so, we first note that we can evaluate the various terms in (\ref{eq:cfunction}) without resorting to any numerical derivatives, as shown in equation (\ref{eq:ds_dl}),
\begin{align}\label{eq:dervas}
  S'(l) = \frac{L_y}{4 G_N} \frac{1}{v_*^2} \qquad , \qquad \mathcal S'(l) = \frac{L_y}{4 G_N}\left(\frac{1}{v_*^2} - \hat \gamma(v_*) Q(v_*)\right) \ .
\end{align}
Recall that $v_*(l)$ is different for the two functionals. It is worth pointing out that the knowledge of the boundary data for these derivatives in a given gauge theory is enough to fix the dual bulk metric within error margin consistent with the statistical noise inherent to measurements \cite{Jokela:2020auu}.\footnote{At zero density, lattice data for entanglement entropy measurements in the case of four-dimensional pure glue $SU(N_c)$, $N_c=2,3,4$, Yang-Mills theory has been extracted in \cite{Buividovich:2008kq,Nakagawa:2011su,Itou:2015cyu,Rabenstein:2018bri}.}
Now, plugging (\ref{eq:dervas}) back into (\ref{eq:cfunction}), we find an expression which admit expansions in various regimes through its dependence of $v_*$ (where $l$ is set to the same value for both the RT and generalized entanglement surfaces). The final expressions for $\mathcal{C}_{UV}$ and $\mathcal{C}_{IR}$, equations (\ref{CUVresult}) and (\ref{CIRresult}), are plotted as a function of $T/\mu$ in the right panel of Fig.~\ref{fig:cfunction}. We observe that both $\mathcal{C}_{UV}\to1$ and $\mathcal{C}_{IR}\to1$ for large enough $T/\mu$. This is because in this regime the AdS-RN solution always dominates over the electron cloud. The dependence of these quantities on $T/\mu$ is, perhaps, more interesting. From (\ref{CUVresult}), and some numerical analysis, we can infer that $\mathcal{C}_{UV}$ is proportional to the ratio between the charges of the outer and inner solutions (times an $\mathcal{O}(1)$ factor), which are interpreted as the total and fractionalized charges, respectively. In other words, we find that
\bea
\mathcal{C}_{UV}&\sim& \frac{Q_{\text{total}}^2}{Q_{\text{BH}}^2}\,,\qquad\,\,\,\,\,\,\, Q_{\text{total}}\equiv Q_{\text{bulk}}+Q_{\text{BH}}\nonumber\\
&\propto& \frac{Q_{\text{cohesive}}^2}{Q_{\text{fractionalized}}^2}\quad (T/\mu\ll1)\,,\label{CUVpropto}
\eea
which means that $\mathcal{C}_{UV}$ can be used to efficiently diagnose and quantify the amount of cohesive charges in the bulk. This dependence is confirmed in Fig.~\ref{fig:cfunction}, in particular, from the fact that $\mathcal{C}_{UV}$ is shown to increase monotonically as $T/\mu$ is decreased. Moreover,
$Q_{\text{bulk}}\to\infty$ as $T/\mu\to0$, so $\mathcal{C}_{UV}$ diverges in this limit as well. Finally, the dependence of $\mathcal{C}_{IR}$ with respect to $T/\mu$ appears to be non-monotonic, which is due to a delicate interplay between the shadows of the electron cloud and RN solutions.

\section{Discussion\label{sec:discussion}}

In this paper we have studied in detail the possibility of detecting charge fractionalization through various information-theoretic probes in strongly coupled gauge theories, using the tools and power of holography. Among the quantities that we analyzed are various entanglement related probes: entanglement entropy, mutual information and entanglement of purification. These quantities provide different measures of correlations across different degrees of freedom and energy scales. Interestingly, since the existence of cohesive charges in the bulk substantially modifies the IR of the theory ({\emph{i.e.}}, the near-horizon region), we find that a detailed characterization of the various measures (in specific corners of the space of parameters) can be used to diagnose the existence of the aforementioned charges.

For instance, the entanglement entropy $S(A)$ for a strip of length $l$ can be used to access the IR region, provided we focus on sufficiently large widths, $l\to\infty$. In this regime, entanglement entropy becomes extensive and turns out to be proportional to the thermal entropy density $S(l\to\infty)\propto s_{\text{th}}(T,\mu)$. This property is particularly useful if we focus on the $T/\mu\to0$ limit. In the absence of cohesive charges, the near-horizon region universally approach an AdS$_2\times$R$^{d-1}$ from which one can deduce that $S(l\to\infty)\propto \mu^{d-1}$ ($d=3$ in our setup). Notice that the finite entropy in this limit indicates a large degeneracy of the ground state. In contrast, the presence of bulk charges induces a large backreaction in the IR. In the $T/\mu\to0$ limit, this leads to an infinitely long throat with a Lifshitz scaling symmetry, from which we can infer that $S(l\to\infty)\propto T^{\frac{d-1}{z}}\mu^{(d-1)(1-1/z)}$, as discussed around equation (\ref{eecloud}). The analysis of mutual information $I(A,B)$ and entanglement of purification $E_P(A,B)$ lead to very similar conclusions. In these two cases, however, the subsystem of interest was taken to be the union of two disconnected strips of length $l$, separated by a distance $s$. The regime of interest in this system was found to be the limit when $l\to\infty$, $s\to0$ and $T/\mu\to0$. In this regime, we also discovered interesting scalings with $T$ and $\mu$ that could be used as a proxy for charge fractionalization, explained around (\ref{MIscaling}) and (\ref{ewcsres}), respectively. We also discussed the possibility of diagnosing the precise position of the electron cloud in the bulk by tracking down jumps in derivatives (of sufficiently high order) of $S(l)$, $I(l,s)$ and $E_P(l,s)$ as a function of $l$ and $s$. However, as we explained in the main text, this is just a feature of the model (which have fluid sources with compact support) and not a property of charge fractionalization \emph{per se}. In more general cases, {\emph{e.g.}}, whenever the bulk charges are distributed smoothly across the bulk, we expect that such kinks would disappear.

We further studied a dynamical probe that characterizes how fast quantum correlations spread in space: the butterfly velocity. This quantity has raised substantial attention in recent years, due to emergent connections between many-body quantum chaos and black hole physics. Previous holographic studies have shown that the butterfly velocity $V_B$ acts as a low-energy Lieb-Robinson velocity $V_{LR}$ which, in the context of quantum information theory, arises as a bound on the rate of transfer of information. It is also known, again through holography, that this quantity provides an upper bound on charge diffusion along black hole horizons; hence, given the system at hand, we expected it to provide us with a useful tool to for diagnosing the existence of cohesive charges in the bulk, in the low temperature regime. Interestingly, our results confirmed our expectations:
we found that this quantity depends on the inner charge parameter $\hat q$ in a particular way, indicated in equation (\ref{VBgeneral}), which turns out to scale very differently with $T$ and $\mu$ in the cloud and pure RN solutions. For instance, in the absence of cohesive charges, one has again a universal (nearly) AdS$_2\times$R$^{d-1}$ geometry at low temperature, from which one can deduce that $V_B\to0$ as $T/\mu\to0$. In this case the bound on charge transport is exact and one finds that the diffusion constant $D_c\to0$, signalizing the dissipative nature of the fractionalized degrees of freedom in the IR. In the presence of the cloud, on the other hand, we find that $V_B$ remains finite in this limit. This in turn indicates the existence of an additional charge sector in the bulk that does not exhibit dissipation, {\emph{i.e.}}, cohesive charges. We also pointed out, and confirmed numerically, that even though the transition between the cloud and the RN solution is of third order, the jump in butterfly velocities across the transition only scales only as the square of the temperature, $\Delta V_B\propto(T_c-T)^2$, which may be easier to track than the jump in free energies.

One quantity that we proposed, worth further highlighting, is the generalized entanglement entropy $\mathcal{S}(A)$, computed holographically through the functional (\ref{eq:functionalM}). The motivation to look for such a functional was partly based on the observation that the bulk surfaces which are used to compute all the previous observables only probe the geometry but are highly insensitive to the presence of bulk charges. At the technical level, we motivated the definition through application of the Iyer-Wald formalism (commonly used in studies of black hole thermodynamics) to a theory of gravity coupled to a $U(1)$ gauge field. The detailed analysis for the derivation of the functional was presented in Appendix~\ref{sec:functional_derivation}. It is worth noticing that, in the context of black hole thermodynamics, this functional is meant to be evaluated at the black hole horizon surface. However, by doing so one finds that the additional term that measures electric flux in the bulk vanishes identically. When evaluated on a different surface, however, this term is generally non-vanishing and, therefore, gives a finite contribution in the type of situations we are interested in. Indeed, we found that one of the main effects of this additional flux term is to repel the bulk surfaces towards the boundary when compared to the RT surfaces, thus, giving rise to a \emph{shadow} region in IR. Based on this observation, we argued that $\mathcal{S}(A)$ must be interpreted as a coarser measure of entanglement entropy for the subsystem $A$ where, besides tracing over all degrees of freedom of the complementary region $A^c$, one also traces over (part) of the fractionalized and coherent charge degrees of freedom contained in $A$. Though, the precise field theoretic definition still remains elusive. An interesting observation is that for small perturbations over AdS (or for small entangling regions in arbitrary excited states), the generalized entanglement entropy defined here is found to obey a first law (\ref{firstlawSgen}) reminiscent of the so-called charged entanglement entropy \cite{Belin:2013uta}, which has a very clear field theoretic replica interpretation. For general excited states, however, the two proposals do not seem to coincide. The reason is that our prescription involves a local chemical potential so, for large enough regions, we generally expect the appearance of a different local weight in the term that measures electric flux.\footnote{We thank Alex Belin for bringing this point to our attention.} Finally, based on the generalized entanglement functional $\mathcal{S}$, we constructed a candidate for a monotonic $c$-function, which we call $\mathcal{C}$, that can be used to efficiently diagnose and characterize the existence of coherent charges across different energy scales. The construction of $\mathcal{S}$ relies on a minor generalization of the standard entropic $c$-function in $(2+1)$-dimensions, $c\propto l^2S'(l)$. Our proposal approximately removes the area term while preserving all of the desired properties for a $c$-function. Furthermore, we showed numerically that our $\mathcal{C}$-function is indeed monotonic in our backgrounds and approaches constant values in the UV and IR, which can be related to the underlying number of charged degrees of freedom in the bulk. To our knowledge, this marks the first time an entropic $c$-function has been shown to meet the criteria akin to Zamolodchikov's theorem at finite chemical potential. In the future, it would be interesting to investigate and understand this function more formally, \emph{i.e.}, from a first principle calculation.

We emphasize that although our analysis was carried out on a particular holographic setup, we expect our analysis and conclusions to hold more generally. We therefore invite more studies in systems where defractionalization occurs at low temperatures and, in particular, in other systems with known gravity duals, either bottom-up or top-down. Of particular interest are the so-called holographic superfluids and superconductors, see, {\emph{e.g.}}, \cite{Hartnoll:2008kx,Horowitz:2008bn,Herzog:2009xv,Gauntlett:2009dn,Ammon:2009xh,Horowitz:2010gk,Bhattacharya:2011tra,Bhaseen:2012gg}. These systems are characterized by the condensation of a charged field in the bulk at sufficiently low temperatures, thus, their corresponding states should exhibit both, cohesive and fractionalized charges when the $U(1)$ symmetry is broken. Another interesting application would be to study situations where both electric and magnetic sources are explicitly present in the bulk \cite{Puletti:2015gwa,Carney:2015dra}. This could yield further insights on, {\emph{e.g.}}, the Haas-van Alphen effect in holographic metals.

Additionally, we believe that further investigations on the general definition and properties of our proposed functional $\mathcal{S}$ and the $\mathcal{C}$-function are in order. For instance, it is not clear what kind of entropic inequalities $\mathcal{S}$ should satisfy, {\emph{e.g.}}, subadditivity, monogamy, etc, or even whether a modified version of these inequalities can be proposed. It would also be interesting to understand the specific field theoretic quantity this functional computes, which might in turn shed light on the mentioned inequalities. Regarding this, we point out that a very recent work \cite{Svesko:2020dfw} studied generalizations of the charged entanglement entropies proposed in \cite{Belin:2013uta}, which seem to be a good starting point for this investigation. We also point out the very nice recent construction in \cite{Zhao:2020qmn} in the context of
Chern-Simons-Einstein gravity, which proposes a simple charged Wilson line prescription that reduces to \cite{Belin:2013uta} for simple states but can be applied to more general excited states. It would be interesting to implement the Iyer-Wald formalism in this setup and compare the resulting functional with their prescription. On the holographic side, it would be useful to understand the role of gauge fields in the semi-classical gravity derivation of holographic entanglement entropy \cite{Lewkowycz:2013nqa}, and try to make contact with the generalized functional proposed here, following the work of Iyer \& Wald. It would also be worthwhile to investigate possible higher derivative corrections to our functional, perhaps, along the lines of \cite{Jacobson:1993xs,Dong:2013qoa,Camps:2013zua}. This would allow us to gauge the interplay between finite 't Hooft coupling corrections and the chemical potential thereof, in systems with charge fractionalization. A further interesting extension would be to come up with an alternative formulation of our functional in terms of bit threads \cite{Freedman:2016zud} using tools of convex optimization \cite{Headrick:2017ucz}. Given the connection between bit threads and entanglement distillation \cite{Agon:2018lwq} (see also \cite{Lin:2020yzf}), this could shed light on the interpretation of the new functional, even in the absence of a concrete boundary dual definition. Finally, one could also ask questions about bulk reconstruction and the emergence of spacetime (either using the generalized entropy $\mathcal{S}$ or the $\mathcal{C}$-function) which have already provided tremendous insights in the program of \emph{gravitation from entanglement} in holography \cite{Lashkari:2013koa,Faulkner:2013ica,Swingle:2014uza,Caceres:2016xjz,Faulkner:2017tkh,Rosso:2020zkk,Agon:2020mvu}.

\paragraph{Acknowledgments:}

It is a pleasure to thank Ulf Gran, Rene Meyer, Christian Northe, Valentina Giangreco M. Puletti, \DJ{}or\dj{}e Radi\v{c}evi\'c, Andrew Svesko, and Suting Zhao for useful discussions and comments on the manuscript, and Alex Belin for collaboration at the initial stages of this work. N.J. is supported in part by Academy of Finland grant no. 1322307. J.F.P. is supported by the Simons Foundation through \emph{It from Qubit: Simons Collaboration on Quantum Fields, Gravity, and Information}. B.S.D. would like to thank Oliver for support and hospitality during the completion of this work.

\appendix

\section{The butterfly velocity for generic backgrounds\label{app:vb}}

In this appendix we will revisit the derivation of the butterfly velocity proposed in \cite{Mezei:2016wfz} and apply it to a general translationally invariant black brane geometry. We will also present the result that is obtained by specializing the formula to a planar RN black hole in AdS, and comment on its interpretation.

Let us start from a $(d+1)-$dimensional metric of the form
\be\label{eq:ansatzmetric}
ds^2=-g_{tt}(v)dt^2+g_{ii}(v)d\vec{x}_{d-1}^2+g_{vv}(v)dv^2\,,
\ee
with boundary at $v=0$ and horizon at $v=v_H$. For our ansatz (\ref{eq:metric}), we can set $d=3$ and
\be
g_{tt}(v)=f(v)\,,\qquad g_{ii}(v)=\frac{1}{v^2}\,,\qquad g_{vv}(v)=g(v)\,,
\ee
but the ansatz (\ref{eq:ansatzmetric}) applies more generally otherwise. Let us now focus on the near-horizon region $v\to v_H$, where the metric functions take the form
\be
g_{tt}(v)\simeq c_0(v_H-v)\,,\quad g_{ii}(v)\simeq g_{ii}(v_H)-g_{ii}'(v_H)(v_H-v)\,,\quad g_{vv}(v)\simeq \frac{c_1}{v_H-v}\,,
\ee
where $c_0$ and $c_1$ are two positive constants and
\be
T=\frac{1}{4\pi}\sqrt{\frac{c_0}{c_1}}\,,
\ee
gives the Hawking temperature. Now, we specialize to Rindler coordinates by replacing
\be
(v_H-v)=\left(2\pi T\right)^2\frac{\rho^2}{c_0}\,.
\ee
With this change, the near-horizon geometry transforms to
\be
ds^2\simeq -\left(2\pi T\right)^2\rho^2dt^2+ \left[g_{ii}(v_H)+\frac{g_{ii}'(v_H)}{g_{tt}'(v_H)}\left(2\pi T\right)^2\rho^2\right]d\vec{x}_{d-1}^2+d\rho^2\,.
\ee
Notice that in the square brackets we have also make the replacement $c_0=-g_{tt}'(v_H)$.

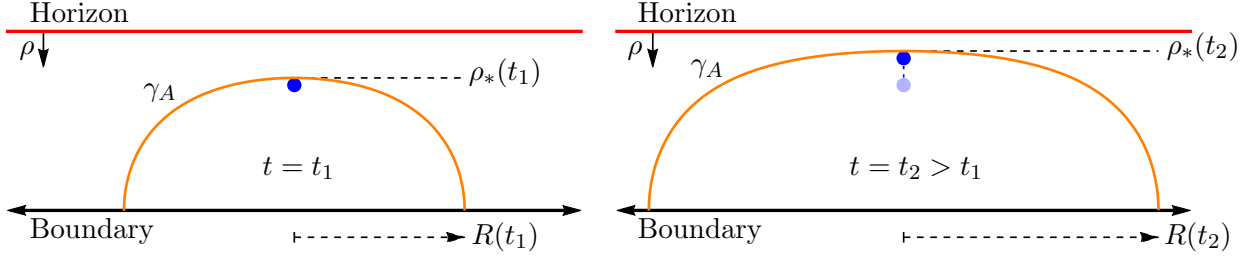
\begin{figure}[t!]
 \centering
     \includegraphics[width=0.48\textwidth]{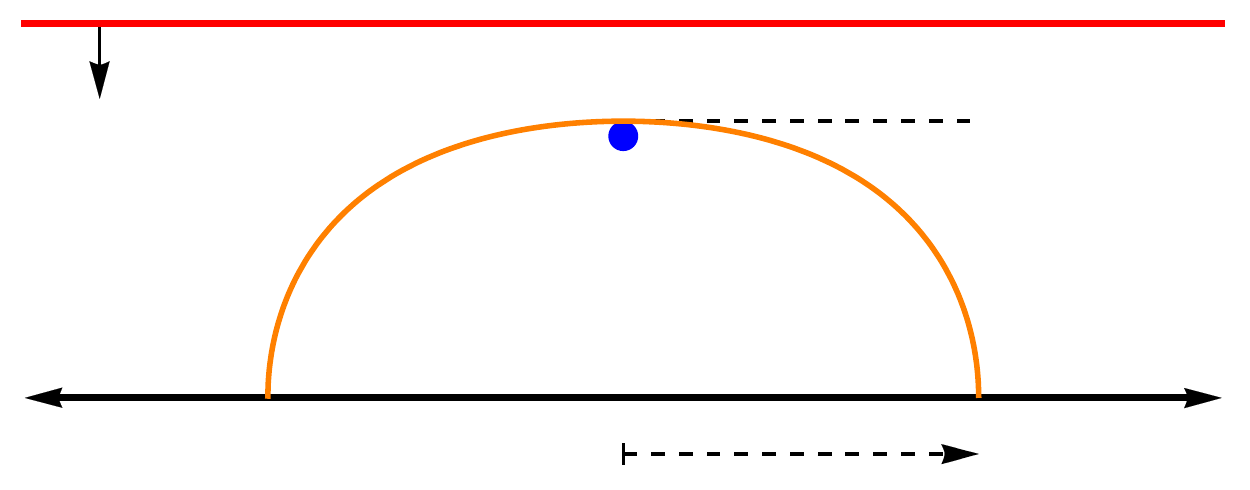}
     \includegraphics[width=0.48\textwidth]{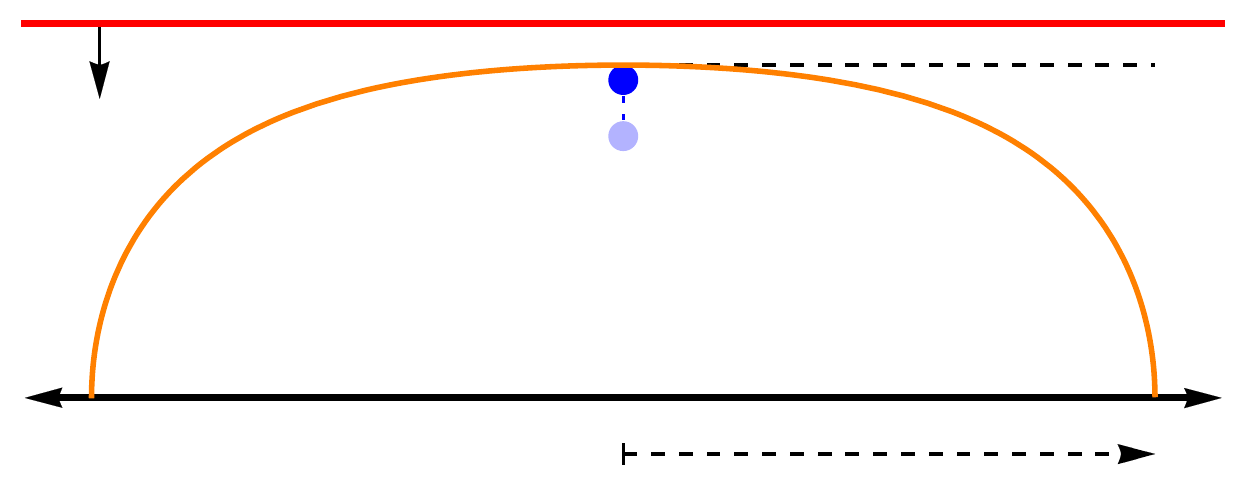}
     \begin{picture}(0,0)
\put(-452,76){\small$ \rho$}
\put(-222,76){\small$ \rho$}
\put(-282,66){\small$ \rho_*(t_1)$}
\put(-18,76){\small$ \rho_*(t_2)$}
\put(-281,4){\small$ R(t_1)$}
\put(-19,4){\small$ R(t_2)$}
\put(-405,60){\small$ \gamma_A$}
\put(-198,69){\small$ \gamma_A$}
\put(-360,30){\small$ t=t_1$}
\put(-138,30){\small$ t=t_2>t_1$}
\put(-448,6){\small Boundary}
\put(-218,6){\small Boundary}
\put(-448,88){\small Horizon}
\put(-218,88){\small Horizon}
\end{picture}
 \caption{\small Schematic diagram that illustrates the computation of the butterfly velocity $V_B$ using concepts of subregion-subregion duality \cite{Mezei:2016wfz}. The infalling particle (depicted in blue) represents a local perturbation that was created near the boundary at early times. The two figures shown correspond to snapshots of the configuration at late times $t_1$ and $t_2$, with $t_2>t_1$. The smallest entanglement wedge that contains the particle at each time determines the function $R(t)$, which at late time can be shown to grow as $R(t)\sim V_B t$.}\label{fig-vb}
   \label{fig:vb}
\end{figure}

Now, following \cite{Mezei:2016wfz} we consider an infalling particle that arises by the insertion of a local operator $V$ in the boundary CFT ({\emph{i.e.}}, a local quench \cite{Nozaki:2013wia,Agon:2020fqs}). At late times, the particle approaches the horizon at a universal (exponential) rate, which in Rindler coordinates is written as
\be\label{rhopart}
\rho(t)=\rho_0 e^{-2\pi T t}\,.
\ee
For chaotic systems, the presence of this excitation leads to the expansion of other non-commuting operators and to a nontrivial commutator squared of the form (\ref{eq-C(t,x)}). The butterfly velocity $V_B$, which characterizes this rate of growth, can be diagnosed by finding the smallest entanglement wedge that contains such particle at late times \cite{Mezei:2016wfz}. See figure \ref{fig-vb} for a pictorial representation.
To do this calculation, we parametrize the RT surface bounding this wedge with a single function $\rho(x^i)$, and pick local coordinates $\xi^i=x^i$. We note that at late times, the RT surfaces that we are interested in sweep the deep IR (near-horizon, or small $\rho$) geometry, and they correspond to large boundary regions. The area functional in this regime reads
\be
\text{Area}(\gamma_A) = g_{ii}(v_H)^{\frac{d-1}{2}} \int d^{d-1}x \left[1+\frac{(d-1)(2 \pi T )^2g_{ii}'(v_H) \rho^2}{2g_{ii}(v_H)g_{tt}'(v_H)} +\frac{(\partial \rho)^2}{2g_{ii}(v_H)} \right].
\ee
Upon minimizing this action we are let to the following equation for the embedding function:
\be
\nabla^2 \rho(x^i) = \nu^2 \rho(x^i)\,,\qquad \nu^2\equiv (d-1)\left(2 \pi T\right)^2 \frac{g_{ii}'(v_H)}{g_{tt}'(v_H)}\,.
\ee
Fortunately, this equation can be solved analytically. The solution is:
\be
\rho(x^i)=\rho_* \frac{\Gamma(n+1)}{2^{-n}\nu^{n}} \frac{I_n(\nu |\vec{x}|)}{|\vec{x}|^n}\,,\qquad n\equiv \frac{d-3}{2}\,,
\ee
where $\rho_*$ denotes the turning point (in Rindler coordinates) and $I_n$ is a modified Bessel function of the second kind. Since these RT surfaces correspond to large boundary regions, when $\rho\gtrsim1/T$ the surface exits the near-horizon and approaches the boundary very fast, almost perpendicularly. Then, we can estimate the size of the region, $R$, in terms of $\rho_*$ by inverting the relation
\be
\frac{1}{T}\simeq\rho_* \frac{\Gamma(n+1)}{2^{-2}\nu^{n}} \frac{I_n(\nu R)}{R^n}\,.
\ee
We note that at large $R$ the above equation simplifies to:
\be\label{rhostar}
\rho_* \simeq e^{-\nu R}\,.
\ee
Comparing (\ref{rhostar}) with (\ref{rhopart}), and requiring that $\rho_* \leq \rho(t)$, {\emph{i.e.}}, that the particle is contained within the entanglement wedge, it follows that
\be
\nu R \geq 2\pi T t\,,
\ee
which implies
\be\label{vbgeneral}
R \geq v_{B} t\,,\qquad  v_{B}\equiv \frac{2\pi T}{\nu}=\sqrt{\frac{g_{tt}'(v_H)}{(d-1)g_{ii}'(v_H)}}\,.
\ee

For example, for $(d+1)-$dimensional AdS black branes we have that
\be
ds^2=\frac{1}{v^2}\left[-F(v)dt^2+\frac{dv^2}{F(v)}+d\vec{x}^2\right]\,,\qquad F(v)=1 -Mv^d\,,
\ee
so
\be
g_{tt}(v)=\frac{F(v)}{v^2}\,,\qquad g_{ii}(v)=\frac{1}{v^2}\,.
\ee
Using (\ref{vbgeneral}) leads to
\be
V_B=\sqrt{\frac{d}{2(d-1)}}\,,
\ee
where the dependence on $M$ completely drops out. For $d=3$, in particular, we have $V_B=\sqrt{3/4}$. For a RN black brane in general dimensions ($d\geq3$) we have
\be
F(v)=1 -Mv^d+Q^2v^{2(d-1)}\,,
\ee
We can define dimensionless coordinates, which is equivalent to rescaling $M$ and $Q$ as
\be
M\to 1+\frac{d-2}{d-1}\hat{q}^2\,,\qquad Q^2\to \frac{d-2}{d-1}\hat{q}^2\,.
\ee
Here $\hat{q}$ ranges from zero to the extremal value, $0\leq \hat{q}\leq \hat{q}_{\text{max}}$, where
\be
\hat{q}_{\text{max}}\equiv \sqrt{\frac{d(d-1)}{(d-2)^2}}\,.
\ee
In this case, we find
\be
V_B=\sqrt{\frac{d}{2(d-1)}\left(1-\frac{(d-2)^2}{d(d-1)}\hat{q}^2\right)}\,.
\ee
Thus, $V_B$ interpolates from the conformal value, in the UV, to zero, in the IR:
\be
V_B(\hat{q}\to0)=\sqrt{\frac{d}{2(d-1)}}\,,\qquad V_B(\hat{q}\to \hat{q}_{\text{max}})\to0\,.
\ee
The fact that $V_B$ vanishes in the extremal limit can be attributed to the fact that all effective degrees of freedom in the IR theory dual to an extremal RN black hole are dissipative.

\section{Iyer--Wald formalism\label{app:IyerWald}}

In this appendix we review basic entries of the Iyer--Wald (or Noether charge) formalism, widely used in the context of black hole thermodynamics. We will start with the basic picture leading to the standard black hole entropy formula and holographic entanglement entropy, and then include the effects of a $U(1)$ gauge field. This will be used to motivate our proposed functional (\ref{eq:functionalM}) for a coarse grained measure of entanglement in situations where the bulk theory contains explicit sources for the $U(1)$ gauge field.

The starting point is a diffeomorphism invariant theory of gravity with Lagrangian:
\begin{equation}
\mathbf{L}=\mathcal{L}(\psi)\mbox{{\boldmath$\epsilon$}}\ ,
\end{equation}
where $\psi$ denotes all dynamical fields and $\mbox{{\boldmath$\epsilon$}}$ is the volume element. The variation of the Lagrangian can be written as follows:
\begin{equation}
\delta\mathbf{L}=\mathbf{E}_{\psi}\delta \psi +d \mathbf{\Theta} \ ,
\end{equation}
where $\mathbf{\Theta}=\mathbf{\Theta}(\psi,\delta\psi)$ is the so-called symplectic potential form
and $\mathbf{E}_{\psi}$ are the equations of motion for the fields.
Now, let $\xi$ be any smooth vector field. One can define the Noether current
as:
\begin{equation}\label{NCcurrent}
\mathbf{J}[\xi]=\mathbf{\Theta}(\psi,\mathcal{L}_{\xi} \psi)-\xi
\cdot \mathbf{L}\ ,
\end{equation}
where $\mathcal{L}_{\xi}$ denotes the Lie
derivative and the dot denotes the contraction of
$\xi^a$ into the first index of the differential form $\mathbf{L}$. A standard calculation leads to:
\begin{equation}
d\mathbf{J}[\xi]=-\mathbf{E}_\psi\mathcal{L}_{\xi}\psi\ .
\end{equation}
This implies that $\mathbf{J}[\xi]$ is closed when the equations of
motion are satisfied. Hence, there must be a form $\mathbf{Q}[\xi]$ such that, whenever
$\psi$ satisfy the equations of motion, we have:
\begin{equation}
\mathbf{J}[\xi]=d\mathbf{Q}[\xi]\ .
\end{equation}
More generally, the Noether charge $\mathbf{Q}[\xi]$ can be
defined in the ``off shell" form so that:
\begin{equation}
\mathbf{J}[\xi]=d\mathbf{Q}[\xi]+\xi^a\mathbf{C}_a\ ,
\end{equation}
where $\mathbf{C}_a$ is locally constructed from all dynamical
fields and vanishes when the equations of motion are satisfied.
In the seminal paper \cite{Wald:1993nt}, Wald showed that for general stationary black holes, black hole entropy may be expressed as an integral
of the Noether charge $\mathbf{Q}[\xi]$ over the horizon:
\begin{equation}
S_{BH}=2\pi \int _{\mathcal{H}} \mathbf{Q}[\xi]\ ,
\end{equation}
where $\xi$ here is the Killing field that generates the bifurcation
surface $\mathcal{H}$.\footnote{The Killing
field has been normalized so that the surface gravity
equals to 1.} Later in \cite{Iyer:1994ys}, Iyer and Wald realized that the
Noether charge can be generally written as:
\be\label{Nchargegen}
\mathbf{Q}[\xi]=\mathbf{W}_a\xi^a+\mathbf{X}^{ab}\nabla_{[a}\xi_{b]}\ ,
\ee
where $(\mathbf{X}^{ab})_{c_3\ldots c_n}\equiv-E_R^{abc_1c_2}\mbox{{\boldmath $\epsilon$}}_{c_1\ldots c_n}$, $E_R^{abcd}$ is the functional derivative of the Lagrangian with respect to the Riemann (with metric held fixed) and
$\mathbf{W}_a$ is locally constructed out of the dynamical fields. Since $\xi$ vanishes at the bifurcation surface, then it follows that the first term of (\ref{Nchargegen})
does not contribute to the black hole entropy. We will ignore this term for now. The second term can be put into a more convenient form by integrating it by parts and evaluating it at the bifurcation surface $\mathcal{H}$, in which case the term $\nabla_{[a}\xi_{b]}=\mbox{{\boldmath $\epsilon$}}_{ab}$ gives the binormal to the
surface. Putting it together, we arrive to an alternative expression for black hole entropy:
\begin{equation}
\label{waldentropy}
S_{BH}=-2\pi\int_{\mathcal{H}}\frac{\p \mathcal{L}}{\p R_{abcd}}\mbox{{\boldmath
$\epsilon$}}_{ab}\mbox{{\boldmath $\epsilon$}}_{cd}\ .
\end{equation}
This formula is purely geometric and does not refer to any vector field $\xi$. Indeed, evaluating it at a different surface $\Gamma_A$
this formula reduces in Einstein gravity to the holographic prescription for computing entanglement entropy in theories with a gravity dual \cite{Ryu:2006bv}:
\be\label{RTfromIW}
S(A)=-2\pi\int_{\Gamma_A}\frac{\p \mathcal{L}}{\p R_{abcd}}\mbox{{\boldmath
$\epsilon$}}_{ab}\mbox{{\boldmath $\epsilon$}}_{cd}=\frac{1}{4G_N}\int_{\Gamma_A}d^{d-1}{\sigma}\sqrt{h}\ ,
\ee
where $\Gamma_A$ is the surface with minimal area subject to appropriate boundary conditions and $h$ is the induced metric on the surface. A few comments are in order. First, notice that we have implicitly assumed that $\Gamma_A$ is a local bifurcation surface on which a bulk Killing vector vanishes. This point can safely be ignored. Since the bulk geometry is locally flat, we can choose to work in the so-called Riemann normal coordinates; this is, we can boost the original metric with appropriate factor at each neighboorhood such that $\Gamma_A$ looks locally like a Rindler horizon \cite{Iyer:1994ys}. Second, notice that for more general theories of gravity, the Wald functional (\ref{waldentropy}) has to be supplemented by extrinsic curvature terms in order to correctly reproduce the universal terms in the entanglement entropy \cite{Dong:2013qoa,Camps:2013zua}. These terms are non-important for the computation of black hole entropy since the extrinsic curvature of the horizon is always zero, but are crucial when evaluating the functional on more general surfaces. We will restrict ourselves to Einstein gravity, so we will ignore such extrinsic curvature corrections. Finally, notice that the first term in (\ref{Nchargegen}) does not necessarily vanish when we evaluate it on an arbitrary surface. The reason is that, depending on $\mathbf{W}_a$, we do not always have the freedom to work in surface-adapted coordinates, {\emph{i.e.}}, we cannot always find a $\xi$ that vanishes at a given surface. We will discuss this point more in detail in the next section and present an explicit example.

\subsection{Gravity coupled to a $U(1)$ gauge field} \label{sec:functional_derivation}
For a diffeomorphism invariant theory of gravity coupled
to a gauge field $\mathbf{A}$, we start with the following Lagrangian:
\be
\mathbf{L}=\mathcal{L}(g_{ab},R_{abcd},A_{a},F_{ab})\bve\ ,
\ee
where $R_{abcd}$ is the Riemann tensor and $F_{ab}$ is the $U(1)$ field strength. Variation of this Lagrangian leads to:
\be
\delta\mathbf{L} = \mathbf{E}_g^{ab}\delta g_{ab}+\mathbf{E}_{A}^{a}\delta A_a+d\mathbf{\Theta}
\ee
where
\bea
&&\mathbf{E}_g^{ab}=\left(\frac{\p \mathcal{L}}{\p g_{ab}}+\frac{1}{2}g^{ab}\mathcal{L}+\frac{\p\mathcal{L}}{\p R_{cdea}}R_{cde}\,\!^{b}+2\nabla_c\nabla_d\frac{\p\mathcal{L}}{\p R_{cabd}}\right)\bve \\
&&\mathbf{E}_{A}^{a}=\left(\frac{\p \mathcal{L}}{\p A_{a}}+2\nabla_b\frac{\p \mathcal{L}}{\p F_{ab}}\right)\bve\ ,
\eea
are the equations of motion and
\be\label{sympAg}
\mathbf{\Theta}_{a_1\cdots a_{n-1}}=\Bigg{(}2\frac{\partial \mathcal{L}}{\partial F_{ab}}\delta
A_{b}
+2\frac{\partial\mathcal{L}}{\partial R_{abcd}}\nabla_d \delta
g_{bc}-2\nabla_{d}\frac{\partial\mathcal{L}}{\partial
R_{dbca}}\delta g_{bc}\Bigg{)}\mbox{{\boldmath
$\epsilon$}}_{aa_1\cdots a_{n-1}}\ .
\ee
is the symplectic potential. For an arbitrary vector field $\xi$, the Lie derivative of $\xi$ on the fields are:
\be
\mathcal{L}_{\xi}g_{ab}=\nabla_{a}\xi_b+\nabla_b\xi_a\,
,\quad\mathcal{L}_{\xi}A_{a}=\nabla_{a}(\xi^b
A_b)+\xi^bF_{ba}\, .
\ee
Substituting these Lie derivatives into (\ref{sympAg}) we arrive at
\bea
&&\mathbf{\Theta}_{a_1\cdots
a_{n-1}}=\Bigg{[}2\nabla_{b}\left(\frac{\partial
\mathcal{L}}{\partial F_{ab}}\xi^c
A_c\right)-\nabla_b\left(\frac{\partial\mathcal{L}}{\partial
R_{abcd}}\nabla_{[c}\xi_{d]}\right)\Bigg{]}\mbox{{\boldmath
$\epsilon$}}_{aa_1\cdots a_{n-1}}\nonumber\\
&&\qquad\qquad\quad\,\,+\Bigg{[}2\frac{\partial \mathcal{L}}{\partial
F_{ab}}\xi^cF_{cb}+2\nabla_{b}\frac{\partial \mathcal{L}}{\partial F_{ab}}\xi^c
A_c\Bigg{]}\mbox{{\boldmath
$\epsilon$}}_{aa_1\cdots a_{n-1}}\ .
\eea
The first line in the above equation will give the Noether charge form, while the second line together with  the terms in $\xi\,\cdot\,
\mathbf{L}$ in (\ref{NCcurrent}) will give the constraints which corresponds to the equations of motion for the metric and gauge field. Thus, we find that
the Noether charge is the sum of two contributions:
\be
\mathbf{Q}=\mathbf{Q}^{g}+\mathbf{Q}^A\ ,
\ee
where
\bea
&&\mathbf{Q}^g_{a_1\cdots a_{n-2}}=-\frac{\partial\mathcal{L}}{\partial R_{abcd}}\nabla_{[c}\xi_{d]}\mbox{{\boldmath $\epsilon$}}_{aba_1\cdots a_{n-2}} \\
&&\mathbf{Q}^A_{a_1\cdots a_{n-2}}=\frac{\partial \mathcal{L}}{\partial F_{ab}}\xi^c A_c\mbox{{\boldmath $\epsilon$}}_{aba_1\cdots a_{n-2}}\ .
\eea
We notice that the $\mathbf{Q}^A$ has exactly the form as the first term of (\ref{Nchargegen}). As we mentioned there, in the context of black hole thermodynamics
such a term does not contribute to the entropy, since $\xi$ vanishes at the horizon. However, if we integrate over a different surface,
this term yields a finite contribution. This observation lead us to define the following functional:
\be\label{eq:functionalgen}
\mathcal{S}(A)=-2\pi \int_{\tilde{\Gamma}_A}\left(\frac{\partial {\cal L}}{\partial R_{\mu\nu\alpha\beta}}\mbox{{\boldmath
$\epsilon$}}_{\mu\nu}\mbox{{\boldmath
$\epsilon$}}_{\alpha\beta}-\gamma\frac{\partial {\cal L}}{\partial F_{\alpha\beta}}\mbox{{\boldmath
$\epsilon$}}_{\alpha\beta} \right)\ ,
\ee
where $\gamma\equiv \xi\cdot A$. The first term is simply the area term of standard holographic entanglement entropy (\ref{RTfromIW}). The second term can be analyzed as follows. First, for a canonically normalized theory of gravity coupled to a gauge field,\footnote{More specifically, we assume an action of the form (with $\kappa^2\equiv8\pi G_N$): $$S=\int d^{d+1}x\sqrt{-g}\left[\frac{1}{2\kappa^2}\left(R-2\Lambda\right)-\frac{1}{4e^2}F^2\right]\,$$} we have that
\be
\frac{\partial {\cal L}}{\partial F_{\alpha\beta}}=-\frac{1}{2e^2}F^{\alpha\beta}\ .
\ee
Next, we use the relations between the field strength and the electric and magnetic vector fields,
\be
E_\mu=u^{\nu}F_{\nu\mu}\,,\qquad B_{\mu}=u^{\nu}\mbox{{\boldmath$\epsilon$}}_{\nu\mu\alpha\beta}F^{\alpha\beta}\ ,
\ee
or alternatively,
\be
F_{\alpha\beta}=(E_\alpha u_\beta-E_\beta u_\alpha)+u^{\mu} \mbox{{\boldmath $\epsilon$}}_{\mu\alpha\beta\nu}B^{\nu}\ ,
\ee
where $u^\mu$ is the time-like future directed unit normal to the time slice $\Sigma_t$, which contains $\tilde{\Gamma}_A$. In the following we will assume a purely electric field strength, in which case the result is proportional to the component of the electric field normal to $\tilde{\Gamma}_A$. The sum of the two terms yields:
\be\label{eq:functional}
\mathcal{S}(A)=\frac{1}{4G_N}\int_{\tilde{\Gamma}_A}d^{d-1}{\sigma}\sqrt{h}\left[1-\frac{\kappa^2}{e^2}\gamma E_\perp\right]\,,\qquad E_\perp \equiv E_\mu n^\mu \ ,
\ee
where $n^{\mu}$ is the outward pointing space-like unit normal to $\tilde{\Gamma}_A\subset\Sigma_t$. A few comments are in order. First, we notice that there is an ambiguity in the choice of Killing vector $\xi$, because in generic situations, we cannot guarantee the existence of a Killing vector that generates the surface $\tilde{\Gamma}_A$; moreover, since we are considering the possibility of having charged matter in the bulk, we cannot always work in Riemann normal coordinates as local boosts would generate currents (and hence magnetic fields) leaving us unable to interpret our surface as a Rindler horizon. The main implication of this observation is that there is generically no choice of $\xi$ for which the second term vanishes. Now, given that we will only be dealing with static black hole solutions, the most natural choice (and possibly the only one) would be to pick the generator of time translations $\xi=\partial_t$ as is done in black hole thermodynamics. In this case $\gamma$ yields the local chemical potential
\be
\gamma=\frac{e}{\kappa}\frac{h(v)}{\sqrt{f(v)}}=\mu_{loc}\ .
\ee
Second, since the gauge field $A$ appears explicitly in the definition of $\gamma$ one might be worried about gauge dependence, in particular, in the possibility of adding a constant to $A$ to make the second term dominate over the first. However, in holography such large gauge transformations are not allowed, because they change the value of the potential at the boundary, and this would imply changing the boundary theory. Third, we note that the above expression has a striking similarity to the Hartnoll-Radi\v{c}evi\'c functional, proposed in \cite{Hartnoll:2012ux}, although in their formula $\gamma$ is taken to be a constant. If we focus on the infrared part of the geometry (as was considered in \cite{Hartnoll:2012ux}), {\emph{i.e.}}, the near horizon region, the local chemical potential becomes approximately constant, and we indeed recover their functional. Our functional thus generalizes their prescription in a very natural way, by weighing the flux term by a local chemical potential. We expect that this generalization will provide a more refined order parameter for charge fractionalization, as explained in the main body of the paper.

\subsection{Linear fluctuations around AdS and a generalized first law}

Let us consider an on-shell perturbation over the vacuum:
\be
g_{\mu\nu}\to g^{(0)}_{\mu\nu}+\delta g_{\mu\nu}\,,\qquad A_\mu\to A^{(0)}_\mu+ \delta A_\mu\ .
\ee
The background metric $g^{(0)}_{\mu\nu}$ is given by pure AdS$_4$ in Poincar\'e coordinates, while $A^{(0)}_\mu$ is an arbitrary constant vector.
We work in Fefferman-Graham coordinates, where the perturbations satisfy $\delta g_{zz}=\delta g_{z\mu}=\delta A_z=0$. Furthermore, from the near-boundary
behavior of $\delta g_{\mu\nu}=z H_{\mu\nu}$ and $\delta A_\mu = z K_\mu$ we can extract the expectation value of the stress tensor and
current dual to the metric and the gauge field, respectively:
\be
\delta\langle T_{\mu\nu}\rangle = \frac{3}{2\kappa^2}H_{\mu\nu}\,,\qquad \delta\langle J_{\mu}\rangle =\frac{1}{e^2}K_\mu\ .
\ee
In this context, it is useful to define the form
\be\label{chi:form}
 \mbox{{\boldmath $\chi$}}=\delta \mathbf{Q}-\xi\cdot\mathbf{\Theta}\ .
\ee
where $\delta \mathbf{Q}$ is the variation of the Noether charge under the on shell perturbation, and $\mathbf{\Theta}$ is the symplectic potential evaluated on this on-shell perturbation. Specializing to the case where $\xi$ is a bifurcate Killing vector field, it can be shown that {\boldmath
$\chi$} is closed
\be
 d\mbox{{\boldmath $\chi$}}=0\ .
\ee
Next, one can make use of Stokes' theorem. Integrating over a spatial slice $\Sigma_t$ between the bifurcation surface $\Gamma_A$ and the boundary region $A$ yields the generalized first law:
\be
\int_{\Sigma_t} d\mbox{{\boldmath $\chi$}}=\int_{\Gamma_A} \mbox{{\boldmath $\chi$}}-\int_A \mbox{{\boldmath $\chi$}}=0 \ .
\ee
Fortunately, for spherical regions in the vacuum we do know a Killing vector $\xi$ that generates the surface $\Gamma_A$, in this case a spherical cap given implicitly by
\be\label{sph:surf}
t=0\,,\qquad x^2+y^2+z^2=R^2\ .
\ee
The Killing vector field that generates this surface is given by:
\be
 \xi=-\frac{2\pi}{R}t(z\partial_z+x^i\partial_i)+\frac{\pi}{R}(R^2-t^2-x^2-y^2-z^2)\partial_t\ .
\ee
Since $\xi$ vanishes at the location of the surface (\ref{sph:surf}), the second term of (\ref{chi:form}) does not contribute to the integral over $\Gamma_A$. In fact, this integral yields
\be
\int_{\Gamma_A} \mbox{{\boldmath $\chi$}}=\delta \int_{\Gamma_A}\mathbf{Q}=\delta \mathcal{S}(A)\ .
\ee
A quick calculation yields
\be\label{bdyint}
\int_A \mbox{{\boldmath $\chi$}}=2\pi\int_Ad^2x\frac{R^2-r^2}{2R}\delta\langle T_{00}\rangle+2\pi\mu\int_Ad^2x\frac{R^2-r^2}{2R}\delta\langle J_0\rangle\ ,
\ee
where $\mu=A^{(0)}_t$ is the chemical potential of the boundary theory. Putting everything together, we find a generalized first law of the form
\be\label{firstlawSgen}
\delta \mathcal{S}(A)=\delta \langle\hat{H}_A\rangle+\mu \,\delta \langle\hat{Q}_A\rangle\ ,
\ee
where we have defined the \emph{modular} charge as the last term appearing in (\ref{bdyint}).\footnote{A similar attempt in deriving a first law of entanglement that includes the contribution of a bulk $U(1)$ field was given in \cite{Hasegawa:2019tai}. In their work, they consider a gauge transformation, equation (3.15) in their paper, which leads to a different first law, equation (3.24). In particular, their charge term does not include the kernel that we obtain in the last term of (\ref{bdyint}).} We note that this is the expected behavior for small variations over the vacuum of the charged entanglement entropies defined in \cite{Belin:2013uta}.\footnote{More generally, one can define charged Renyi entropies as: $$ S_A^n(\mu_E)= \frac{1}{1-n} \log \text{Tr}\left( \tilde{\rho}_A(\mu_E) \right)^n\,, \qquad \tilde{\rho}_A(\mu_E)\equiv \rho_A e^{\mu_{E} A} \,.$$
To obtain the charged entanglement entropy one could set $\mu_E = \mu(n-1)$ and look at the limit $n\to1$.}
In general excited states, however, the two proposals do not coincide. The reason is that our prescription involves a local chemical potential so, for large enough regions, we generally expect the appearance of a different local weight in the flux term.

\section{Generalized functional on a disk\label{app:disk}}

In the main text the only boundary subregions we considered were strips. We find the same qualitative features also for disks which are defined as the region $x_1^2+x_2^2\leq R^2$, where $R$ is the radius of the disk. Due to rotational symmetry it is convenient to write the spatial boundary directions in the metric using polar coordinates
\begin{align}
  d\hat x_1^2 + d\hat x_2^2 = d\hat r^2 + \hat r^2 d\hat\phi^2 \ .
\end{align}
The profile of the bulk surface is given by $\hat r = \hat r(\hat v)$. The generalized functional in this case is
\begin{align}
  4 G_N \mathcal S = 2\pi \int \left( \frac{\hat r(\hat v)}{\hat v^2} \sqrt{\hat g(\hat v) \hat v^2 + \hat r'(\hat v)^2} + \hat\gamma(\hat v) Q(\hat v)  \hat r(\hat v) \hat r'(\hat v) \right) d\hat v \ . \label{eq:generalized_disk_entropy}
\end{align}
This time there is no cyclic coordinates so we must solve the profile $\hat r(\hat v)$ from the full Euler-Lagrange equations
\begin{align}
  \frac{d}{d\hat v} \frac{\partial \mathcal L}{\partial \hat r'(\hat v)} - \frac{\partial \mathcal L}{\partial \hat r(\hat v)} = 0 \ ,
\end{align}
where $\mathcal L$ is the integrand of \eqref{eq:generalized_disk_entropy}. Like with the strip, there is a point $\hat v = \hat v_*$ past which the bulk surface $\hat r(\hat v)$ does not extend. The relationship between $R$ and $\hat v_*$ is such that $R = 0$ corresponds to $\hat v_* = 0$ and when $R$ is increased, $\hat v_*$ increases monotonously.

The minimal surfaces of boundary disks behave in a way analogous to the strips of Sec.~\ref{sec:generalized_functional}. If we compare the minimal surfaces of \eqref{eq:generalized_disk_entropy} with RT-surfaces anchored to the same boundary region, we find that the RT-surfaces reach deeper into the bulk. There also exists a point $\hat v_* = \hat v_s$ where the corresponding boundary disk size $R$ diverges. In other words, boundary anchored minimal surfaces of \eqref{eq:generalized_disk_entropy} can not probe the bulk past some $\hat v_s < \hat v < 1$. Furthermore, we find that this shadow region lies at the same point $\hat v_s$ as it did when we considered boundary strips. This leads us to conjecture that the shadow is a feature of the background independent of the shape of the boundary subregion. The IR-behavior of \eqref{eq:generalized_disk_entropy} is determined by $\hat v_s$ and our background functions
\begin{align}
  \left .4 G_N v_H^2 \mathcal S\right|_{R\to\infty} = \left( \frac{1}{\hat v^2} - \hat\gamma(\hat v_s) Q(\hat v_s) \right) \pi R^2 + \mathcal O (R) \ .
\end{align}

\section{$\mathcal{C}$-function expansions for the entangling strip\label{app:cfunctions}}
In this appendix we derive analytic expressions for the proposed $\mathcal{C}$-function (\ref{eq:cfunction}) in various limits of interest. Before diving into the calculation, it will be useful to rewrite (\ref{eq:cfunction}) in a more convenient way. To do so, we note that our $\cal C$-function involves first derivatives of the entanglement and generalized entanglement entropies, $S'(l)$ and $\mathcal S'(l)$, respectively. Following \cite{Jokela:2020auu} (see also \cite{Lowenstein:2021asp}) we now derive convenient expressions for such derivatives.

For the ease of notation, we start by writing the two functionals in the following form:
\begin{align}
  F = 2 \int_\epsilon^{v_*} \mathcal L(x'(v), v) dv \ ,
\end{align}
where $\epsilon$ is the UV-cutoff, $v_*$ is the turning point, and $x'(v)$ is the profile of the corresponding minimal surface. We now make the variation $x(v) \to x(v) + \delta x(v)$, while keeping the boundary conditions fixed. In particular, we require that $\delta x(v_*) = 0$ so that the surface is connected at the tip. Then, the variation of the functional yields
\begin{align}
  \delta F = 2 \int_\epsilon^{v_*} \frac{\partial \mathcal L}{\partial x'} \delta x'(v) dv = 2 \int_\epsilon^{v_*} \frac{d}{d v} \left( \frac{\partial \mathcal L}{\partial x'} \delta x \right) dv - 2 \int_\epsilon^{v_*} \frac{d}{d v} \frac{\partial \mathcal L}{\partial x'} \delta x dv \ .
\end{align}
The last integral is zero as a consequence of the equations of motion so
\begin{align}\label{eq:vari}
  \delta F = 2 \frac{\partial \mathcal L}{\partial x'}(v_*) \delta x(v_*) - 2 \frac{\partial \mathcal L}{\partial x'}(\epsilon) \delta x(\epsilon) = - \frac{\partial \mathcal L}{\partial x'} \delta l \ .
\end{align}
Here we have used the fact that $\delta x(v_*) = 0$ and $\delta x(\epsilon) = \delta l / 2$ where $l$ denotes the width of the boundary strip. Also, note that $\partial \mathcal L / \partial x'$ is a constant along the minimal surface, since it is nothing but the conserved momentum associated with the shift symmetry $x\to x+\rm{constant}$.
We may now evaluate (\ref{eq:vari}) at any point along the minimal surface. For convenience, we choose to evaluate it at the tip $v=v_*$, in which case one finds
\begin{align}
  \frac{\delta F}{\delta l} = - \frac{\partial \mathcal L}{\partial x'} (v_*) \ .
\end{align}
This formula is intuitive, since the variation of the functional with respect to $l$ naturally yields the momentum.\footnote{Similar arguments can be made for holographic Wilson loops, where one can additionally interpret the conserved momentum as a force acting on the external quark traversing the plasma \cite{Gutiez:2020sxg}.} However, the momentum does not need to be a constant when the tip value is not fixed in the transverse directions, {\emph{e.g.}}, for disk entangling regions.
Finally, specializing to the functionals at hand, $S(l)$ and $\mathcal S(l)$, we find
\begin{align}
  \frac{4 G_N}{L_y L^2} \frac{d S}{d l} = \frac{1}{v_*^2}\,, \qquad \text{and} \qquad \frac{4 G_N}{L_y L^2} \frac{ d\mathcal S}{d l} = \frac{1}{v_*^2} - \hat \gamma(v_*) Q(v_*) \,. \label{eq:ds_dl}
\end{align}
As a consistency check, note that these expressions match the conserved momenta associated with the RT functional (\ref{eq:conserv}) and generalized functional (\ref{eq:turning_point}), evaluated at the tip.

With the above expressions at hand, we are now ready to derive UV and IR expansions of the $\mathcal{C}$--function presented in Section \ref{sec:c-func}, for the case of the entangling strip.

\subsection{$\mathcal{C}_{UV}$}

For small entangling regions, all bulk surfaces stay very close to the AdS boundary. The calculation in this case boils down to three steps:
\begin{enumerate}
\item Derive the near-boundary expansions for $l_{RT}(\hat v_*)$ and $l_{\mathcal{S}}(\hat v_*)$ as an expansion in $\hat v_*$, where $l_{RT}$ and $l_{\mathcal{S}}$ are given by \eqref{eq:strip_width} and \eqref{eq:generalized_strip_width}, respectively.
\item Invert the expansions from the previous step to find $\hat v_*(l_{RT})$ and $\hat v_*(l_{\mathcal{S}})$.
\item Plug the $\hat v_*$ expansions into the expression for $\mathcal{C}$, using (\ref{eq:ds_dl}).
\end{enumerate}

We recall that between the AdS boundary and the outer edge of the cloud, the metric and gauge field are given by an AdS-RN black brane solution \eqref{eq:exterior_soluion}. We can, therefore, leverage simple analytic functions in our derivation of the near-boundary behavior of $\mathcal{C}$. To simplify our analysis, we start by rewriting our metric and gauge field ansatz as:
  \begin{eqnarray}
    \hat f &=& \frac{\alpha^2}{\hat v^2}\left(1-f_1 \hat v^3+f_2 \hat v^4\right)\,,\label{eq:cuv_f}\\
    \hat g &=& \frac{\alpha^2}{\hat v^4 \hat f} = \frac{1}{\hat v^2(1-f_1 \hat v^3+f_2\hat v^4)}\,,\label{eq:cuv_g}\\
    \hat h &=& \alpha\left(h_0 - h_1 \hat v\right)\,,
  \end{eqnarray}
with
  \begin{eqnarray*}
    f_1 &\equiv& \alpha^{-2}M\,,~~f_2 \equiv \alpha^{-2}Q^2\,,\\
    h_0 &\equiv& \alpha^{-1}\mu\,,~~~h_1\equiv\alpha^{-1}\sqrt{2}Q ~~\big(\!= \sqrt{2 f_2}\,\big)\,.
  \end{eqnarray*}
This allows us to unify our descriptions of the electron cloud and AdS-RN systems. To recover a particular case, we make the substitutions:
\begin{equation}\label{eq:uv-param-replacements}
  \begin{aligned}
    \text{AdS-RN:} &~~~ \alpha\rightarrow 1\,, ~~ M\rightarrow \left(1+\frac{\hat q^2}{2}\right)\,, ~~ Q^2\rightarrow \frac{\hat q^2}{2}\,, ~~ \mu \rightarrow \hat q\,,\\
    \text{EC:} &~~~ \alpha\rightarrow c_s\,,  ~M\rightarrow m_{s}\,, ~~ Q^2 \rightarrow \frac{q_s^2}{2}\,, ~~ \mu\rightarrow \mu_s \,.
  \end{aligned}
\end{equation}
With these definitions in mind, we can write \eqref{eq:strip_width} as:
\begin{equation}
\frac{l_{RT}}{v_H} = 2 \hat v_*\int_0^{1}dx\frac{x^2}{\sqrt{1-x^4}}\left(\frac{1}{\sqrt{1-f_1 \hat v_*^3 x^3 + f_2 \hat v_*^4 x^4}}\right)\,. \label{eq:lrtuv_1}
\end{equation}
Near the AdS boundary, conformal invariance guarantees $\hat v_* \sim l$, and so it makes sense to treat $\hat v_*$ as a small expansion parameter. We can, therefore, evaluate \eqref{eq:lrtuv_1} by expanding the integrand around $\hat v_* = 0$ and integrating the resulting expression term by term:
\begin{equation}
  \frac{l_{RT}}{v_H} = 2\hat v_*\int_0^1dx\frac{x^2}{\sqrt{1-x^4}}\left[1+\frac{f_1\hat v_*^3}{2} x^3-\frac{f_2 \hat v_*^4}{2}x^4+\ldots \right]\,,\label{eq:lrtuv_2}
\end{equation}
where every term can be evaluated using the identity:
\begin{equation*}
  \int_0^1dx~x^{\mu-1}(1-x^{\lambda})^{\nu-1} = \frac{1}{\lambda}\frac{\Gamma\left(\frac{\mu}{\lambda}\right)\Gamma\left(\nu\right)}{\Gamma\left(\frac{\mu}{\lambda}+\nu\right)}\,.
\end{equation*}
Carrying on in this fashion, one finds that the full expression for \eqref{eq:lrtuv_2} can be written as the double sum:\footnote{With some obvious redefinitions, it can be shown that this expression coincides with the double expansion developed in \cite{Kundu:2016dyk}.}
\begin{equation}
  \frac{l_{RT}}{v_H} = \frac{1}{2}\sum_{k=0}^{n}\sum_{j=0}^{k}\frac{ \Gamma \left(k+\frac{1}{2}\right)  \Gamma \left(\frac{1}{4} (j+3 k+3)\right)}{\Gamma (j+1)
  \Gamma (1+k-j) \Gamma \left(\frac{1}{4} (j+3 k+5)\right)}(-f_2)^j f_1^{k-j}\hat v_*^{1+3 k+j}\,.\label{eq:lrtuv_3}
\end{equation}
The near-boundary expansion of $l_{\mathcal{S}}$ can be derived using the same logic as in the RT case. Upon expanding \eqref{eq:generalized_strip_width} around $\hat v_* = 0$ and integrating term by term, one finds:
\begin{equation}
  \frac{l_{\mathcal{S}}}{v_H} = \sum_{k=0}^{n}\sum_{j=0}^{k}\sum_{l=0}^{j}\sum_{m=0}^{l} \left(A^{(k-j),(j-l)}_{(l-m),(l+m)}\right) f_1^{k-j}f_2^{j-l}h_0^{l-m}h_1^{l+m} \hat v_*^{1+3k+j-l+m}\,, \label{eq:lguv_1}
\end{equation}
where $A^{pq}_{rs}$ are numerical coefficients.\footnote{$A^{pq}_{rs}$ does not admit an immediately obvious closed form expression. Finding one is left as an exercise for the curious reader.} The first few coefficients are:
\begin{eqnarray*}
\mathcal{O}(\hat v_*)  &:& A^{00}_{00} = 2\sqrt{\pi}\frac{\Gamma(3/4)}{\Gamma(1/4)}\,,\\
\mathcal{O}(\hat v_*^4)&:& A^{00}_{11} = \sqrt{\pi}\frac{\Gamma(3/4)}{\Gamma(1/4)} -1\,, \qquad A^{10}_{00} = \frac{\pi}{8}\,,\\
\mathcal{O}(\hat v_*^5)&:& A^{00}_{02} = \frac{1}{4\sqrt{2\pi}}\left(\Gamma\left(1/4\right)^2-4\Gamma\left(3/4\right)^2\right)\,,
\qquad A^{01}_{00} = -\frac{3}{5\sqrt{2\pi}}\Gamma\left(3/4\right)^2\,,
\end{eqnarray*}
so the first few terms of \eqref{eq:lguv_1} are:
\begin{eqnarray}
  \frac{l_{\mathcal{S}}}{v_H} &=& 2 \sqrt{\pi }\frac{ \Gamma
  (3/4)}{\Gamma (1/4)}\hat v_{*}+
   \left(\frac{\pi  }{8}f_{1}+ \left(\sqrt{\pi }\frac{ \Gamma (3/4)}{\Gamma
  (1/4)}-1\right)h_{0} h_{1}\right)\hat v_{*}^4\nonumber\\
  &~&+\frac{ \left(5 h_{1}^2 \Gamma (1/4)^2-4 \Gamma
  (3/4)^2 \left(3 f_{2}+5 h_{1}^2\right)\right)}{20 \sqrt{2 \pi }}\hat v_{*}^5 + \mathcal{O}(\hat v_*^7)\,.\label{eq:lguv_2}
\end{eqnarray}

We are now in a position to invert \eqref{eq:lrtuv_3} and \eqref{eq:lguv_1} for $\hat v_*(l_{RT})$ and $\hat v_*(l_{\mathcal{S}})$, respectively, and use them to determine $\mathcal{C}_{UV}$. Substituting \eqref{eq:ds_dl} into \eqref{eq:cfunction}, we find that
\begin{equation}
\mathcal{C} = \frac{\Delta S_{EC}}{\Delta S_{RN}}\,,\qquad\Delta S \equiv \left(\frac{1}{\hat v_*^{2}} - \hat{\gamma}Q\right)_{l_{\mathcal{S}}} - \left.\frac{1}{\hat v_*^2}\right|_{l_{RT}}\,.\label{eq:cstrip}
\end{equation}
Applying the appropriate substitutions for the EC and AdS backgrounds \eqref{eq:uv-param-replacements}, we find:
\begin{equation}
\mathcal{C}= \frac{q_s\mu_s}{c_s^2 \hat q^2}\left\{1 + \left(1 - \frac{q_s}{\mu_s}\right)\left[\left(2 \left(\frac{l}{v_H}\right) \frac{\Gamma(5/4)^2}{\Gamma(3/4)^2} \right)+\left(2 \left(\frac{l}{v_H}\right) \frac{\Gamma(5/4)^2}{\Gamma(3/4)^2} \right)^2  \right] \right\} + \mathcal{O}(l^3)\,,\label{eq:cuv-strip-final}
\end{equation}
hence
\be\label{CUVresult}
\mathcal{C}_{UV} \equiv \mathcal C(l\to0)=\frac{q_s\mu_s}{c_s^2 \hat q^2}\,.
\ee

\subsection{$\mathcal{C}_{IR}$}
For large entangling regions, all surfaces contributing to \eqref{eq:cfunction} are well approximated by rectangular surfaces which lie at $\hat v= \left(\hat v_*\right)_{\text{max}}$ for a length $l$ and then climb straight to the boundary. Using (\ref{eq:ds_dl}), it is easy to see that in this limit we obtain
\begin{align}
\frac{4 G_N v_H^2 S(l)}{L_y}  &= \frac{2 v_H^2}{\epsilon} + l \,,\\
\frac{4 G_N v_H^2 S_{RN}(l)}{L_y} &= \frac{2 v_H^2}{\epsilon} + l \,,\\
\frac{4 G_N v_H^2 \mathcal S(l)}{L_y} &= \frac{2 v_H^2}{\epsilon} + \left( \frac{1}{\hat v_s^2} - \hat\gamma(\hat v_s) Q(\hat v_s) \right) l \,,\\
\frac{4 G_N v_H^2 \mathcal S_{RN}(l)}{L_y} &= \frac{2 v_H^2}{\epsilon} + \left( \frac{1}{\hat v_{s,RN}^2} - \hat\gamma_{RN}(\hat v_{s,RN}) \hat q \right) l \ ,
\end{align}
where $\hat v_s$ and $\hat v_{s,RN}$ are the shadow positions for the electron cloud and AdS-RN geometries, respectively. Plugging the above into \eqref{eq:cfunction} we obtain
\begin{align}\label{CIRresult}
  \mathcal{C}_{IR} \equiv \mathcal C(l\to\infty) = \frac{\hat v_s^{-2} - \hat\gamma(\hat v_s) Q(\hat v_s)-1}{\hat v_{s,RN}^{-2} - \hat\gamma_{RN}(\hat v_{s,RN}) \hat q-1} \ .
\end{align}
We show the behavior of $\mathcal{C}_{IR}$ as a function of $T/\mu$ in Fig.~\ref{fig:cfunction}.

\bibliographystyle{JHEP}
\bibliography{refs}

\end{document}